%% file: LaplaceReview-arXiv.tex
\documentclass{apa}
\usepackage[cmex10]{amsmath}%
\usepackage{amssymb,amsfonts,trsym}
\usepackage{multirow}
\usepackage{graphicx}
\begin{document}
\input{tcmmacros}

\title{Cognitive computation using neural representations of time and space in
the Laplace domain} 

\shorttitle{Cognitive computation using representations in the Laplace domain} 
\leftheader{Cognitive computation using representations in the Laplace domain} 
\rightheader{Howard and Hasselmo} 
\author{Marc W.~Howard and Michael E.~Hasselmo}
\affiliation{Center for Memory and Brain, Center for Systems
Neuroscience, Department of Psychological and Brain Sciences, Department of
Physics\\Boston University}

\abstract{
 Memory for the past makes use of a record of what happened when---a function
 over past time.  Time cells in the hippocampus and temporal context cells in the
 entorhinal cortex both code for  events as a function of past time, but with
 very different receptive fields.  Time cells in the hippocampus can be
 understood as a compressed estimate of events as a function of the past.
 Temporal context cells in the entorhinal cortex can be understood as the
 Laplace transform of that function, respectively.  Other functional cell
 types in the hippocampus and related regions, including border cells, place
 cells, trajectory coding, splitter cells, can be understood as coding for
 functions over space or past movements or their Laplace transforms.  More
 abstract quantities, like distance in an abstract conceptual space or
 numerosity could also be mapped onto populations of neurons coding for the
 Laplace transform of functions over those variables.  Quantitative cognitive
 models of memory and evidence accumulation can also be specified in this framework
 allowing constraints from both behavior and neurophysiology.   More
 generally, the computational power of the Laplace domain could be important
 for efficiently implementing data-independent operators, which could serve as
 a basis for neural models of a very broad range of cognitive computations.  
} 

 \note{Submitted to \emph{Computational Brain \& Behavior} \today}

\maketitle{}

 Connectionist models have had astounding success in recent years in
 describing increasingly sophisticated behaviors using a
 large number of simple processing elements \cite{LeCuEtal15,GravEtal14}.
 However, the native ability to perform symbolic computations has long been
 noted as a key problem in developing a theory of cognition
 \cite{FodoPyly88,GallKing11,Marc18a}.  Among other things, symbolic processing
 requires operators that are independent of the data on which they operate.
 For instance, a computer program can add any pair of integers whether they
 are familiar or not. 
 Human cognition also has a powerful symbolic capability that allows us to
 perform many data-independent operations.  To
 take a concrete situation, after focusing on Figure~\ref{fig:Laplace} one
 could close one's eyes and implement a huge number of operations on the
 contents of memory. For instance, one could choose to imagine Moe Howard's
 face translated by 5~cm to the left.  Or decide if the
 thought bubble in \textbf{A} is  above or below Moe's tie.
 Operations like translation (e.g., imagining
 Moe's face moved by 5~cm to the left) or subtraction (e.g., the relative
 position of the thought bubble and Moe's tie)  would have obvious benefits in
 computational problems like spatial navigation, where we
 have learned a great deal about functional correlates of neurons in the
 hippocampus and entorhinal cortex \cite{OKeeDost71,WilsMcNa93,HaftEtal05}.
 If cognitive data of many different types used the same form of neural
 representation then if we knew how to build data-independent operators in one
 domain, the same computational mechanisms could be reused across many domains
 of cognition.  A complete set of operations would constitute a ``cognitive
 map'' that could be used for many different types of information
 \cite{OKeeNade78,BehrEtal18}. 
 

 
 This paper reviews recent evidence that suggests a common form of neural
 representation for many types of information important in cognition. The
 basic idea is that the firing rate of populations of neurons represent
 functions out in the world.  Some populations do not represent these
 functions directly, but rather represent the Laplace transform of functions.
 Because we know a great deal about the properties of the Laplace transform,
 this lets us understand the computational capabilities of these populations
 at a relatively deep level.  This paper proceeds in three sections.  In the
 section entitled ``Computing with functions in the Laplace domain'' we sketch
 out in non-technical language the ideas behind this hypothesis.  This section
 will explain what it means to say the brain ``represents a function,''  and
 what it means for the a population of neurons to estimate ``the Laplace
 transform of a function.'' In the second section, we describe recent
 neurophysiological evidence from the hippocampus and entorhinal cortex.   The
 data show evidence that hippocampal time cells behave as if they are
 estimating a function over past time.  Moreover neurons in the entorhinal
 cortex behave as if they were estimating the Laplace transform of this
 function over past time.  To the extent one accepts this empirical account,
 it means that the brain has a transform/inverse pair for functions
 of time one synapse away in the medial temporal lobe.  In the third section,
 we will review modeling work describing how to construct transform/inverse
 pairs to represent functions over not only time, but spatial position,
 other kinematic variables, and accumulated evidence for use in
 decision-making circuits.   We suggest that the reader should seriously
 consider  the idea that the brain might use transform/inverse pairs to
 perform cognitive computations in many different domains.  

 \section*{Computing with functions in the Laplace domain}

 \begin{figure}
		 \centering
		 \begin{tabular}{lll}
					\textbf{A} &  
					\textbf{B} &  
					\textbf{C} \\
					 
		 			\includegraphics[height=0.15\textheight]{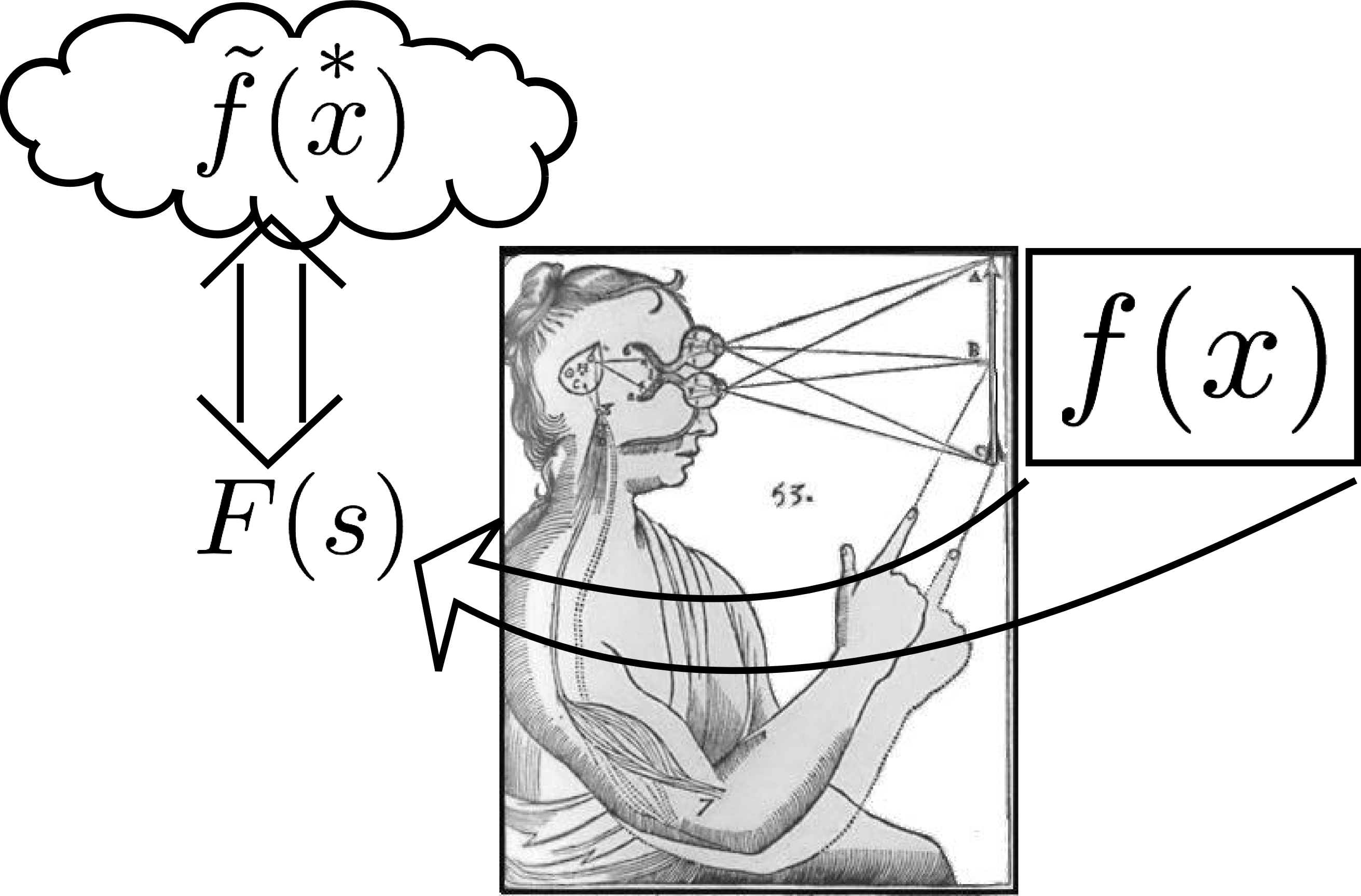}
					&
					\includegraphics[height=0.15\textheight]{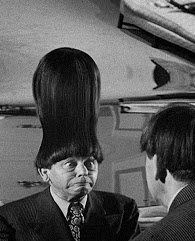}
					& 
					\includegraphics[height=0.2\textheight]{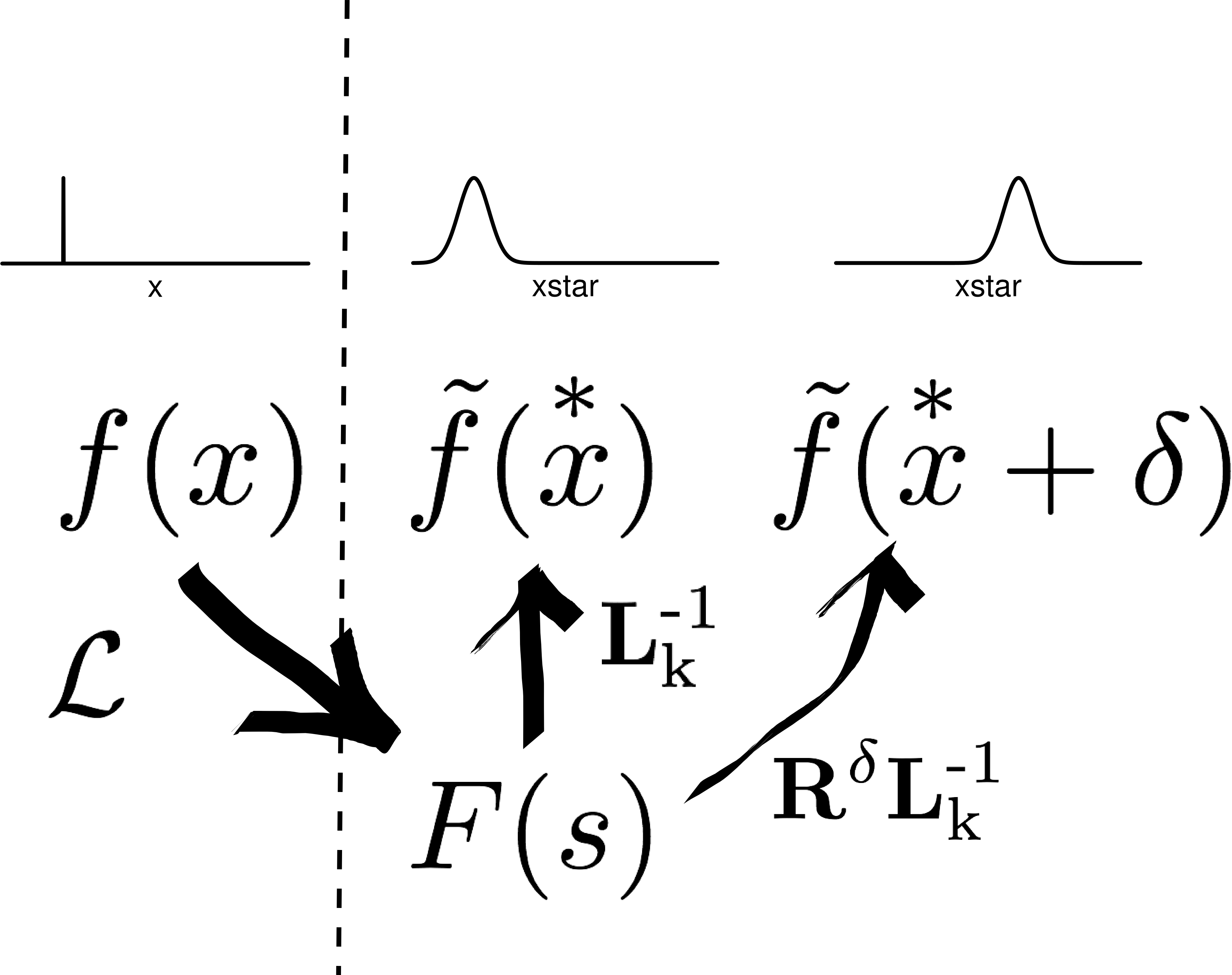}
		 \end{tabular}
		 \caption{\textbf{Encoding functions in the Laplace domain.}
		  \textbf{A.} The brain tries to estimate functions $f(x)$ out in the
		  world.  The brain's estimate of this function is denoted
		  $\ftilde(\xstar)$.  In many cases, it is not practical to directly
		  compute $\ftilde(\xstar)$.  Instead the brain first estimates the Laplace
		  transform of $f(x)$, $F(s)$ and then constructs $\ftilde(\xstar)$ by
		  inverting the transform \emph{via} an inverse transform operator $\Lk$.
		  Both $\ftilde(\xstar)$ and $F(s)$ correspond to firing rate across
		  many neurons indexed by $\xstar$ or $s$ as appropriate.  We assume
		  the population is very large so we can think of $\xstar$ and $s$ as
		  effectively continuous.
		  \textbf{B.} The Laplace transform of a function is analogous to a
		  reflection in a fun-house mirror.  Like the reflection,  the transform of a
		  function need not superficially resemble the original image.
		  However, each unique image causes a unique reflection.  This means
		  that given a particular reflection, and knowledge of the distortion
		  introduced by the mirror, one could reconstruct the particular
		  image associated with the particular reflection.   Similarly, each 
		  function specifies a unique transform, one can in principle
		  reconstruct the original function from its transform.
		  \textbf{C.} 
		  Data-independent operators to compute with functions in the Laplace
		  domain.  The world, left of the dashed line, contains some function,
		  $f(x)$.  The Laplace operator $\mathcal{L}$, is used to generate the
		  Laplace transform of the function $F(s)$ in the brain.
		  Approximately inverting the transform, \emph{via} an operator $\Lk$
		  generates an internal estimate of the external function,
		  $\ftilde(\xstar)$.  Note that there is some ``blur'' in this
		  estimate of the true function.  Data-independent operators are
		  necessary for symbolic computation.  Many such operators can be
		  efficiently implemented in the Laplace domain.  Here we illustrate a
		  translation operator.  Although the world has provided $f(x)$, we
		  want to compute a translated version of the function $f(x+\delta)$.
		  We can rcompute the Laplace transform of $f(x+\delta)$ by operating
		  on $F(s)$  with an operator $\mathbf{R}^\delta$ such that
		  $\mathbf{R}^\delta F(s)$ is the transform of $f(x+\delta)$.  Now,
		  applying the inverse operator we can  obtain an approximation of
		  $f(x+\delta)$, $\ftilde(\xstar+\delta) =
		  \Lk \mathbf{R}^\delta F(s)$.  Note that the translation operator
		  $\mathbf{R}$ is independent of the data---it works equally well on
		  any function.  Moreover, in the case of translation, $\mathbf{R}$ is
		  particularly simple---just a diagonal matrix---enabling efficient
		  computation of translation.   Other data-independent operators have
		  a simple form in the Laplace domain.
 		 }
		 \label{fig:Laplace}
 \end{figure}

 We argue that the brain, at least in some cases, computes using
 \emph{functions} describing information over some continuum
 (Figure~\ref{fig:Laplace}a).  Consider some function $f$  defining a scalar
 value in the external world over some domain $x$, $f(x)$, for instance, in
 vision the pattern of light in a greyscale image as a function
 of retinal position.
 We will write $f(x_o)$ to refer to the value at a single position and 
 understand $f(x)$ to mean the brightness over all possible positions.
 The activity of neurons along the retina along the retinal surface
 estimates this function.  To distinguish the brain's
 internal estimate from the actual function in the world, we will write
 $\ftilde(\xstar)$ to describe the activity over a population of neurons.
 The value at a particular location $\ftilde(\xstar_o)$ corresponds to the
 activation of the receptor that is indexed to the physical location $x_o$. We
 understand $\ftilde(\xstar)$ to mean the activation of all the
 receptors over all their locations.
 The continuous parameters associated with each neuron, $\xstar$ maps onto the
 continuum of $x$, enabling the population to distinguish many different
 functions $f(x)$.  We can understand the particular shape of the receptive
 fields as basis functions over the domain $x$.  
 We will assume that the number of receptors is very large
 and the distance between their centers is small so that we can think of
 $\ftilde(\xstar)$ as if it was a function over a continuous variable.  Note
 that the density of receptors need not be constant in different regions of $x$.    

 If we cannot directly place a receptor at a particular physical location
 $x_o$, how can we estimate functions over variables such as time or
 allocentric position, or location within an abstract conceptual space?  We
 hypothesize \cite{ShanHowa10,ShanHowa12,HowaEtal14} that as an intermediate
 step in estimating $\ftilde(\xstar)$ the brain could construct the Laplace
 transform of $f(x)$ over another population of neurons.  We describe this situation
 notationally as $F(s) = \mathcal{L} f(x)$.  Analogous to the way in which
 $\ftilde(\xstar)$ corresponds to the activity of many neurons indexed by
 their value of $\xstar$,  $F(s)$ is understandable as a particular pattern of
 activity over a population of neurons, each indexed by a continuous parameter
 $s$.  Rather than receptive fields that tile $x$, neurons in $F(s)$ have
 receptive fields that fall off exponentially like $e^{-sx}$.
 
 
 The insight that $F(s)$ is the Laplace transform of $f(x)$ is very
 powerful---it means that knowing $F(s)$ is enough to specify $f(x)$. 
 Because neurons in $F(s)$ do not have receptive fields centered on a
 particular value of $x$, it is not necessarily intuitive to visualize the
 connection between $f$ and $F$.  In this sense, the Laplace transform of a
 function is something like the reflection in a funhouse mirror
 (Fig.~\ref{fig:Laplace}B).  We can construct $\ftilde(\xstar)$ by inverting
 the transform: $\ftilde(\xstar) = \Lk F(s)$.  Here $\Lk$ is a feedforward
 operator that approximates the inverse Laplace transform
 \cite{ShanHowa12,LiuEtal19}.   Of course the inverse cannot be precise---with
 a finite number of neurons we cannot reconstruct the potentially infinite
 amount of information in a continuous function (Appendix~2).  However, it can be
 shown   \cite{ShanHowa12} that the properties of $\Lk$ blur $\ftilde$ such
 that  the width of each receptive field in $\ftilde$ is a constant fraction
 of $\xstar$.  This is closely analogous to the finding that the size of
 receptive fields in the visual system grows proportional to the distance from
 the fovea.


 One of the reasons the Laplace transform is so widely used in engineering and
 data processing applications is because one can efficiently implement
 data-independent operations on functions in the Laplace domain.  That is,
 suppose one wants to perform an operation on some function $f$.   In many
 cases, it is is more computationally efficient computational to construct $F
 = \mathcal{L}f$, apply the appropriate operator in the Laplace domain to $F$
 and then take the inverse to get the desired answer.
 Figure~\ref{fig:Laplace}C provides a schematic for how this could work for
 function translation---constructing $f(x + \delta)$ from $f(x)$.  Efficient
 Laplace domain methods exist for many unary operators that take in one
 function $f$, such as translation, computing the mean or moments of the
 distribution or taking derivatives.  Moreover, there are also methods for
 binary operations that compare two functions $f$ and $g$ to one another, such
 as convolution and cross-correlation.  Thus, if the brain had access to both
 Laplace transforms it could in principle take advantage of some of this
 computational power to implement data-independent operations.  


 \section*{Coding for past events as a function of time in the brain}

\begin{figure}
	\centering
	\begin{tabular}{lclclc}
			\textbf{A} && \textbf{B}\\
			&
			\begin{minipage}{0.35\textwidth}
					\includegraphics[height=0.18\textheight]{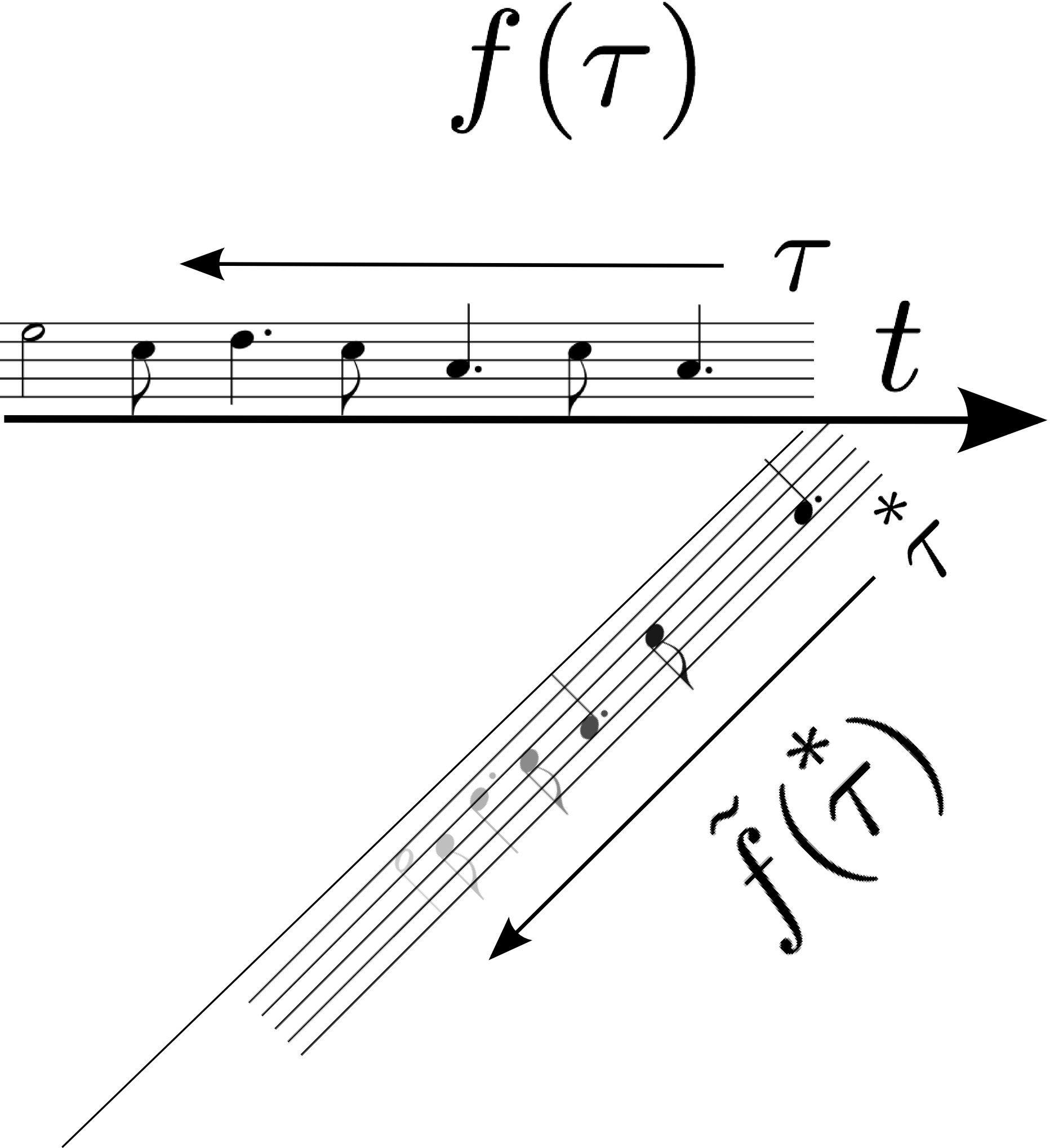}
			\end{minipage}
			&& 
		 	\begin{minipage}{0.35\textwidth}
	        	\includegraphics[height=0.18\textheight]{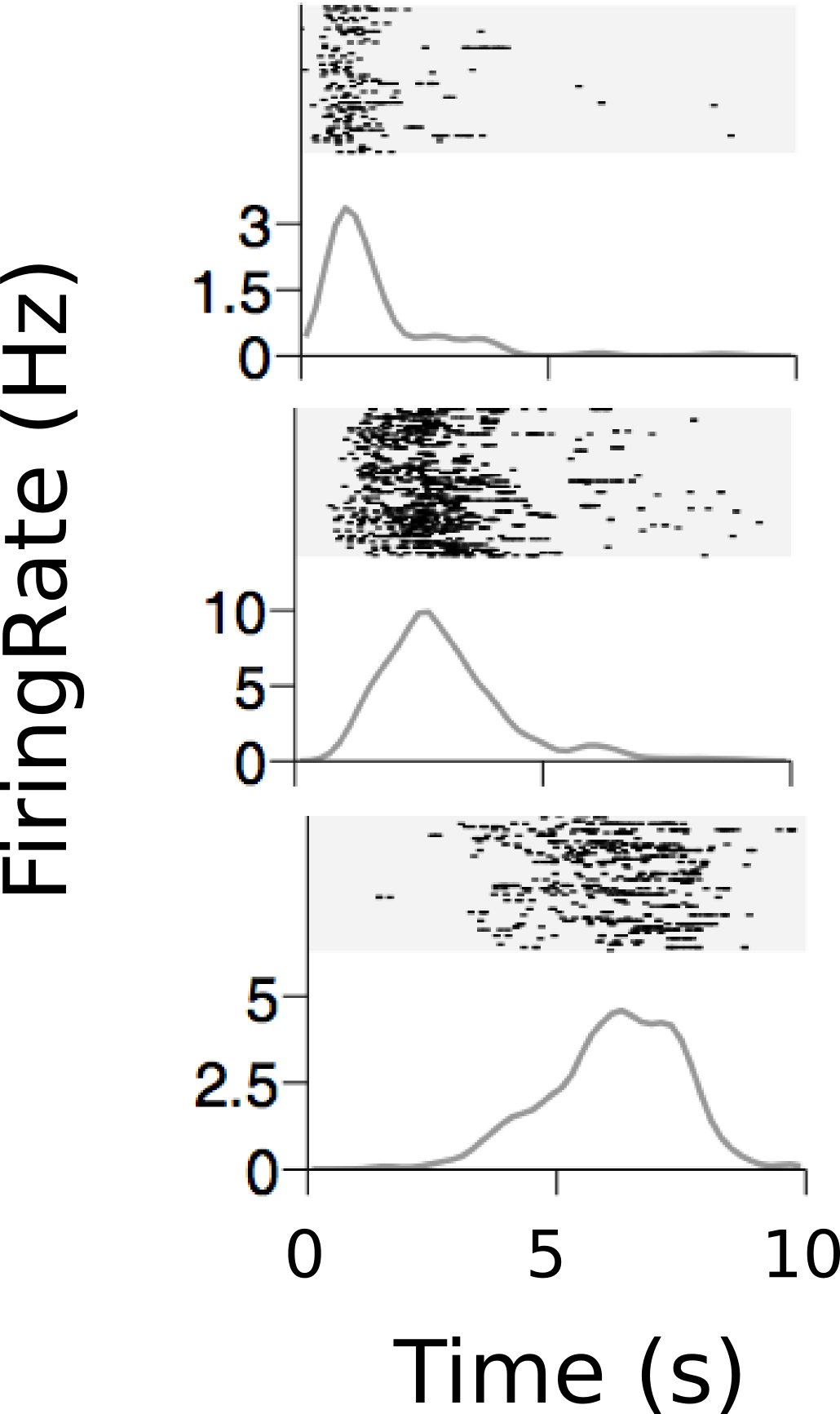}
			\end{minipage}\\
	 \textbf{C} \\
	 &
	 \multicolumn{3}{c}{
			\begin{minipage}{0.7\textwidth}
				\begin{tabular}{lll}
					\includegraphics[height=0.3\textheight]{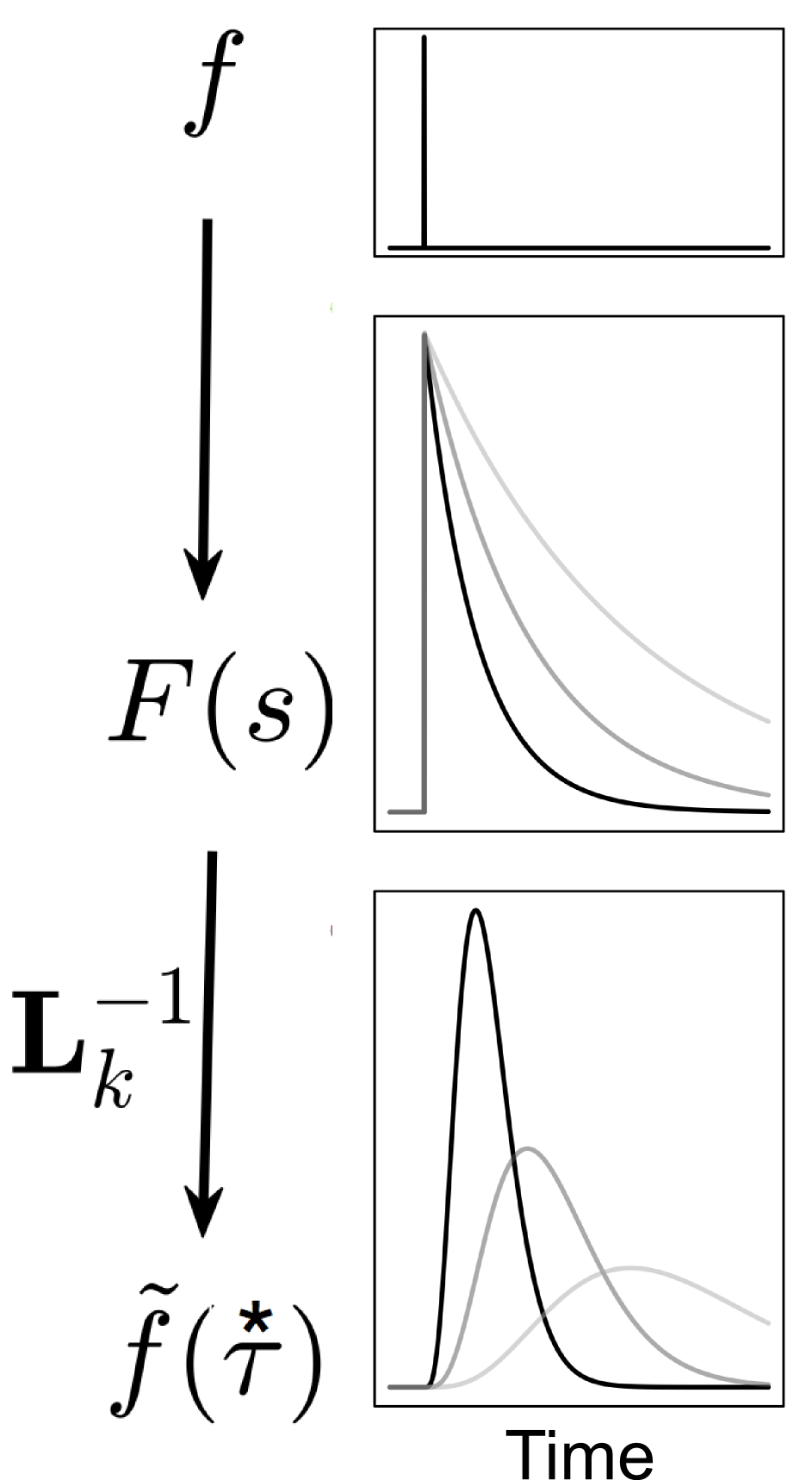}              &
					\includegraphics[height=0.3\textheight]{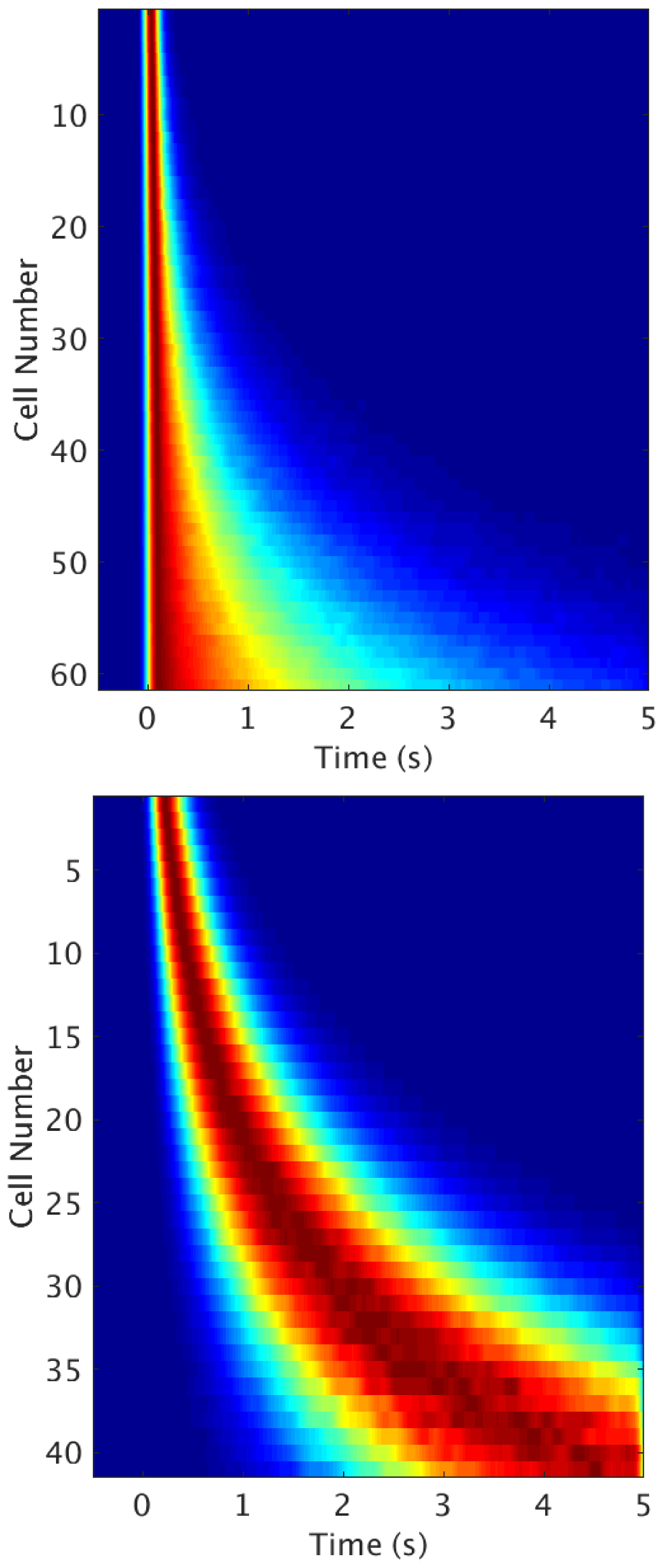}
					& 
				   \includegraphics[height=0.3\textheight]{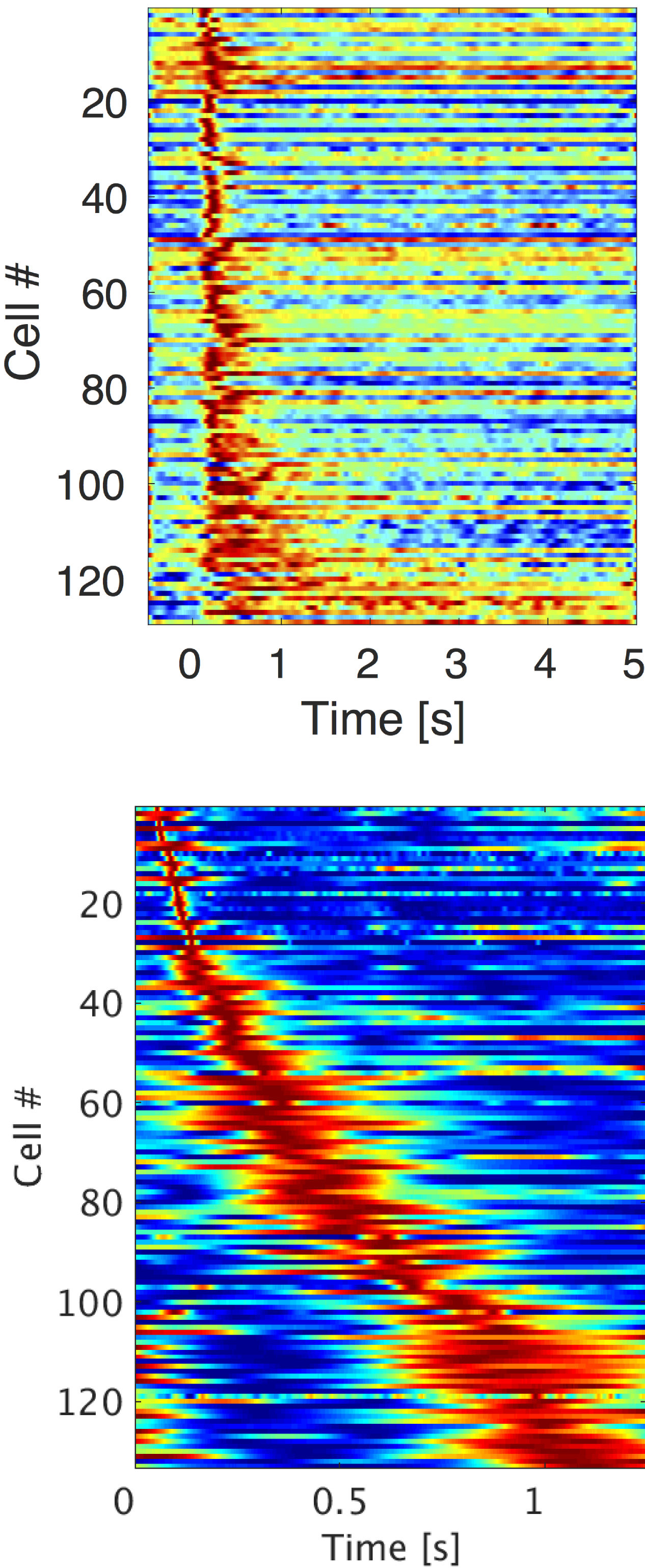}
			   \end{tabular}
			\end{minipage}
			}
	\end{tabular}
	\caption{\small
		\textbf{A compressed  timeline and its Laplace transform in
		hippocampus and entorhinal cortex.}
		\textbf{A.}  Schematic for a compressed internal timeline.  The
		horizontal line describes a sequence of distinct events,  here a
		sequence of tones, in the external world.  At a particular moment $t$, 
		$f(\tau)$ describes the objective past leading up to the
		present, with $\tau=0$ corresponding to the immediate past. 
		At any moment $f(\tau)$ describes what event (i.e., which
		note) happened at each time $\tau$ in the past.  One can imagine an internal
		estimate of the timeline leading up to the present $\ftilde(\taustar)$
		(diagonal line).  $\ftilde(\taustar)$ estimates what happened when in
		the past, but the internal time axis is compressed.  This means that
		the time of occurence for past events is resolved with
		decreasing accuracy for events further in the past (note that spacing
		between the memory for notes further in the past is decreased).  
		\textbf{B.} Hippocampal time cells have receptive fields in time.
		Each panel is a different neuron, with a series of rasters at the top
		and a smoothed peri-stimulus time histogram shown below. Time zero in
		this study is the beginning of the delay period of a memory
		experiment.  Time cells fire when the triggering event is a
		certain time in the past.  They can thus be understood as coding
		for a function over past time.
		Note that the cells that fire later have wider temporal receptive
		fields.  This is general characteristic of hippocampal time cells and
		indicates less temporal resolution for events further in the past,
		consistent with a compressed representation in the brain.
		After MacDonald, et al., (2011).  
		\textbf{C.} Left: Time courses for model units encoding the
		instantaneous input $f$, the Laplace transform of the past $F(s)$, and
		the inverse transform $\ftilde(\taustar)$.  When the input was a time
		$\tau$ in the past, the neurons coding the Laplace transform are
		activated as $e^{-s\tau}$.  Neurons in $F(s)$ activate at the same
		time after the input is presented and then decay exponentially at
		different rates in time as $\tau$ increases.  Neurons coding for
		$\ftilde(\taustar)$ activate  sequentially.
		Middle, Theoretical predictions for the two populations expressed as a
		heatmap.  Right; Empirical heatmaps for units in macaque entorhinal
		cortex (top) and hippocampus (bottom). Data from Bright, et al.,
		(2019) and Cruzado, et al., (2019).  
		\label{fig:Timecells}
	}
\end{figure}

 Memory, by definition, requires some record of the past.  Psychologists have
 long appreciated that memory relies on an explicit record of what events
 happened when in the past \cite{Jame90,BrowEtal07a,BalsGall09}.  
 Computational neuroscientists have long proposed models with
 sequentially-activated neurons could represent past events
 \cite{TankHopf87,GrosMerr92,Gold09}.  The observation that memory
 is less precise for less recent events has led to the proposal that this
 record of the past is compressed, such that the time at which recent events
 occurred has better resolution than events further in the past
 (Fig.~\ref{fig:Timecells}A).  This compression is analogous to the
 compression of the visual system where regions of visual space near the fovea
 have a much greater resolution than regions further from the fovea
 \cite{Howa18}. 

 More formally, at time $t$ the brain tries to estimate the objective past
 leading up to the present $f(\tau)$.  In this formulation, $\tau$ runs from
 zero to infinity with zero corresponding to the moment in the immediate past
 at time $t$.  At each moment, we can understand the past as a function over
 the $\tau$ axis  (Fig.~\ref{fig:Timecells}A).  
 This function $f(\tau)$ is estimated by a population of neurons that we write
 as $\ftilde(\taustar)$.  Cognitive modeling and theoretical work
 \cite{ShanHowa12,HowaEtal15} has shown that this kind of representation can
 be used to construct detailed behavioral models of many 
 memory tasks if the representation of the past is compressed.    
 We first argue that hippocampal time cells have the
 properties predicted for $\ftilde(\taustar)$ and then review evidence
 suggesting that neurons in the entorhinal cortex of rodents and monkeys show
 properties consistent with the Laplace transform $F(s) = \mathcal{L}
 f(\tau)$.  

 \subsection*{Time cells in the hippocampus code for a compressed timeline of
 the recent past}
 Time cells in the hippocampus behave as if they have receptive fields
 organized in time (Figure~\ref{fig:Laplace}C,
 \cite{PastEtal08,MacDEtal11,Eich14,TeraEtal17,TaxiEtal18,CruzEtal19}).   
 As a triggering event recedes into the past, the event first enters and then
 exits the ``time field'' of different time cells indexed by $\taustar$.
 Because the time fields for different cells are centered on different
 $\taustar$s, the population fires in sequence as the triggering event moves
 through the past.   Hippocampal time cells have the computational properties
 one would expect of a compressed representation of what happened when as a
 function of past time.  First, different external stimuli can trigger
 distinct sequences of hippocampal time cells
 \cite{MacDEtal11,TeraEtal17,TaxiEtal18,CruzEtal19},  meaning that these
 populations carry information about what stimulus happened in the past.
 Second, hippocampal time cells show decreasing temporal accuracy further in
 the past.  The number of cells with receptive fields around a particular
 value  $\taustar_o$ goes down as $\taustar_o$ goes up.  Moreover, the
 width of receptive fields go up with $\taustar_o$
 \cite{CruzEtal19,KrauEtal13,HowaEtal14,SalzEtal16}.

 \subsection*{Temporal context cells in entorhinal cortex code for the Laplace
 transform of a compressed timeline of the past.}

 Let us consider how we would identify neurons coding for the Laplace
 transform $F(s) = \mathcal{L} f(\tau)$.  Cells coding the Laplace transform
 of a variable $x$ should show receptive fields that fall off like $e^{-sx}$.
 A set of neurons coding the Laplace transform of past time $\tau$ should show
 receptive fields that go like $e^{-s\tau}$, with many different values of $s$
 across different neurons.  If we think of the triggering stimulus as a delta
 function at time $t=0$, it enters $f(\tau)$ at time $t$ at $\tau=0$.  At time
 $t$ after the triggering stimulus, the firing rates should change like
 $e^{-st}$.  Observing the firing of a neuron with rate constant $s$, we
 should see it change shortly after the triggering stimulus, and then relax
 back to baseline exponentially in the time after the triggering stimulus.
 Cells with high values of $s$ (corresponding to fast time constants) should
 relax quickly; cells with small values of $s$ corresponding to slow time
 constants) should relax more slowly.  We would expect a continuum of $s$
 values to describe the continuum of $\tau$ values.  If the representation 
 is compressed, we would see more neurons with fast decay rates than with slow
 decay rates.  The grey lines in Figure~\ref{fig:Timecells}C depicts how
 $F(s)$ and $\ftilde(\taustar)$ should behave in the time after a triggering
 stimulus for different values of $s$ and $\taustar$.  

 Recent evidence shows that cells in the entorhinal cortex contain temporal
 information, like hippocampal time cells, but with temporal receptive fields
 that are as we would expect from the Laplace transform
 (Figure~\ref{fig:Timecells}C, \cite{BrigEtal19}). 
 These ``temporal context cells'' are analogous to findings from a rodent
 experiment recording from lateral entorhinal cortex \cite{TsaoEtal18}.   In
 that study, neurons in the EC were perturbed by entry into an enclosure for a
 period of random foraging.  Different neurons relaxed with a variety of
 rates, showing gradual decay over time scales of up to tens of minutes.
 Although there are thus far only two studies showing this phenomenon, the
 similarity of the qualitative properties of the neurons despite drastic
 changes in the methods of the two studies is striking.   Appendix~1 discusses
 possible neurophysiological mechanisms to implement the Laplace transform and
 $\Lk$ in neural circuits.

 
 \subsection*{Time and memory outside the MTL}

 The entorhinal cortex and hippocampus are believed to be important in
 episodic memory. Computational modeling suggests that a representation like
 $\ftilde(\taustar)$ is also useful for other ``kinds'' of memory, including
 short-term working memory tasks, conditioning tasks, as well as interval
 timing tasks \cite{HowaEtal15,TigaEtal19}.  This suggests that other brain
 regions have access to representations like $\ftilde(\taustar)$.   Indeed,
 time cells with more or less the same properties of hippocampal time cells
 have been observed in the striatum \cite{MellEtal15,AkhlEtal16,JinEtal09}
 medial prefrontal cortex \cite{TigaEtal17}, lateral prefrontal cortex
 \cite{TigaEtal18a,CruzEtal19} and dorsolateral prefrontal cortex
 \cite{JinEtal09}.  The fact that this kind of representation is so widespread
 suggests that many different types of memory utilize a compressed timeline of
 the past.  Ramping neurons observed outside of the EC during memory and
 timing tasks (e.g., \cite{MitaEtal09,RossEtal19,ZhanEtal19,WangEtal18}) could
 also be manifestations of the Laplace transform of the past, but this hypothesis
 has not thus far been explicitly tested.

\section*{Compressed functions of other variables}
 
\begin{figure}
	\centering
    \begin{tabular}{lll} 
		\textbf{A}& \textbf{B}\\
         \begin{minipage}{0.2\textwidth}
        	\includegraphics[width=0.95\textwidth]{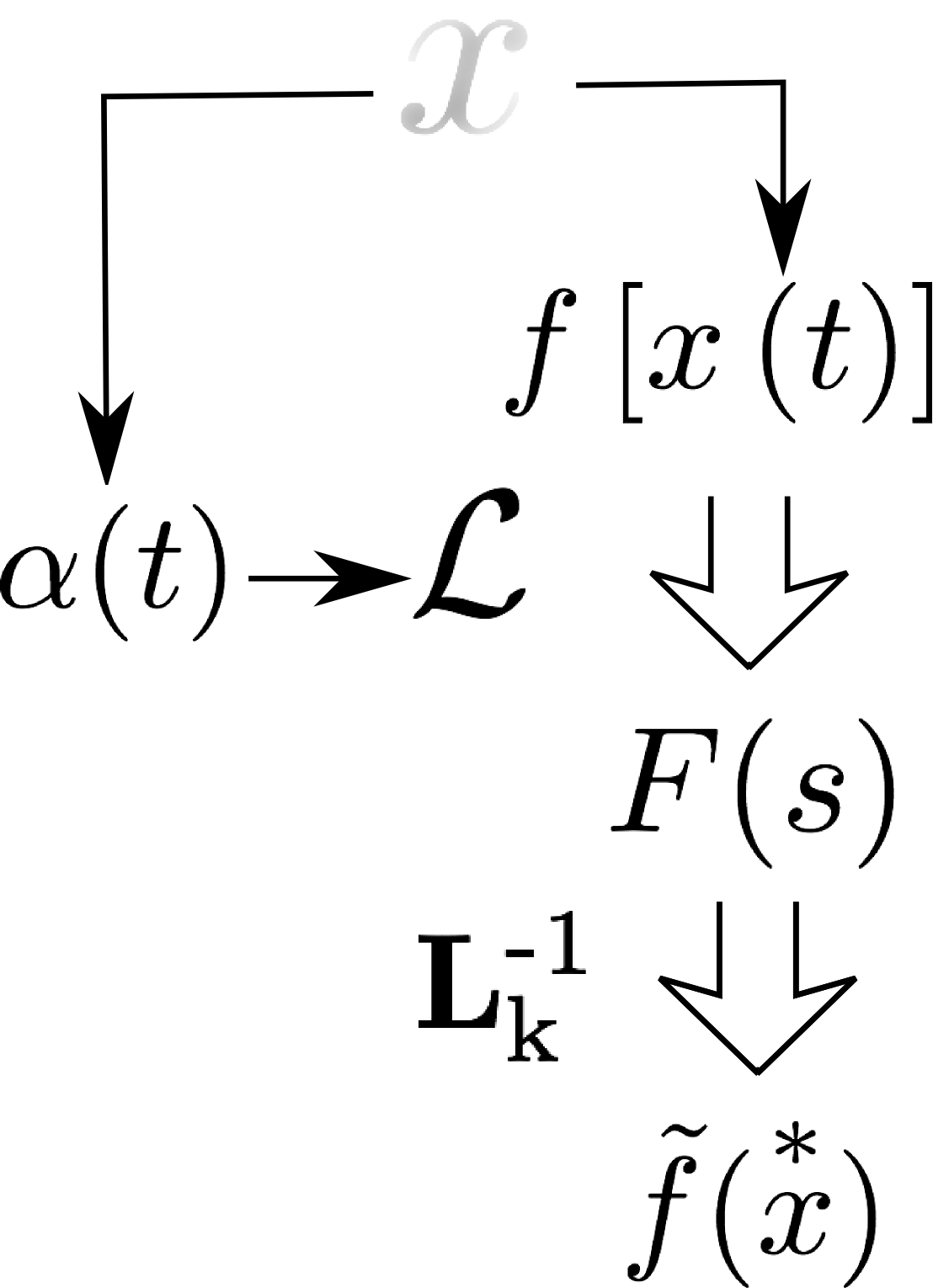}
        \end{minipage}
       &
        \begin{minipage}{0.7\textwidth}
        	\includegraphics[width=0.95\textwidth]{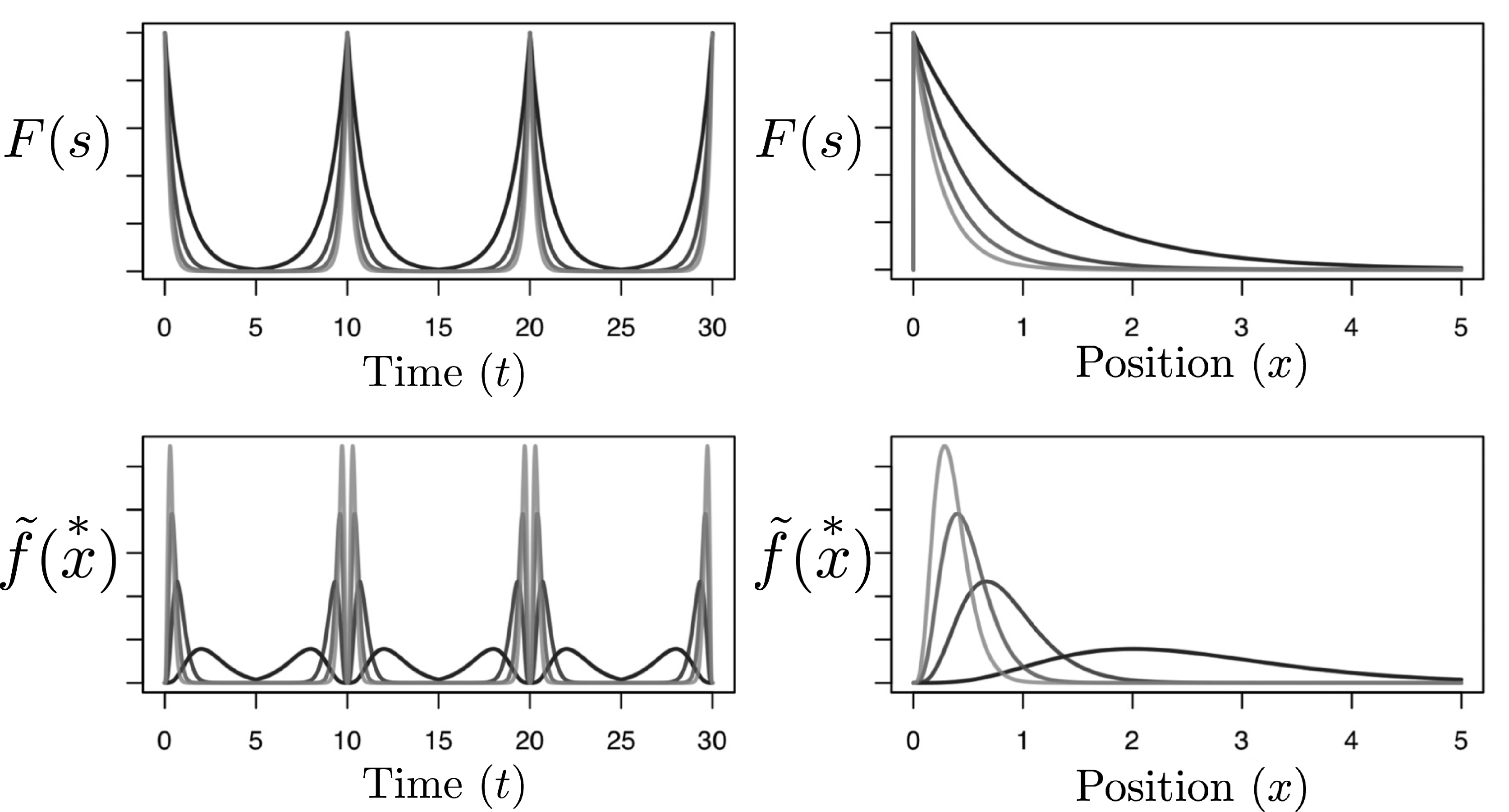}
        \end{minipage}\\
   	\end{tabular}
	\caption{\small
		\textbf{Laplace domain code for space and other variables.} 
		\textbf{A.}  This computational framework can be used to construct
		compressed functions over any continuous variable, here denoted $x$,
		for which the brain has access to the time derivative.  The gain of
		the neurons coding for the transform is dynamically set to $\alpha(t)$
		(see Eq.~\ref{eq:alpha} in Appendix~2). If $\alpha(t) = dx/dt$, the time
		derivative of $x$, then the transform is with respect to $x$ instead
		of $t$.  The inverse thus estimates $f(x)$ rather than $f(t)$.   
		\textbf{B.} Schematic showing the activity of a population of cells
		coding for one-dimensional position from an environmental boundary.
		In this simulation, the landmark is at the left of a linear track
		(position zero).  The animal moves at a constant speed to the other end of the
		track and is reflected back towards its initial starting point.  The
		activity of populations of cells coding the Laplace transform ($F(s)$,
		top row) and the inverse transform ($\ftilde(\xstar)$, bottom) are
		shown as a function of time (left) and position (right) over three
		laps.
		Different values of $s$ and $\xstar$ are shown as different lines.  As
		the animal moves away from the landmark, firing rates in the transform
		decay exponentially with different rates.  When the animal reverses
		direction, these cells rise until the starting position is reached. 
		The cells coding the inverse fire in sequence as the distance to the
		landmark grows and then fire in the reverse sequence as the agent
		approaches the landmark.  When plotted as a function of position
		rather than time the cells in $F(s)$ show characteristic exponential
		receptive fields as a function of position and the cells in
		$\ftilde(\xstar)$ show circumscribed place fields.
 \label{fig:CBB}
	}
\end{figure}

 A general framework for cognitive computation in the brain requires that
 representations of many different variables use the same ``neural
 currency.'' The same formalism utilizing the Laplace transform and its
 inverse can give rise not only to functions over time but functions over many
 other variables as well.  The basic idea (Appendix~2) is the equations
 implementing the Laplace transform of a function of time are  modulated by
 the rate of change of some variable $x$.  We refer to the modulation factor
 at time $t$ as $\alpha(t)$.  At the cellular level, $\alpha(t)$ is
 understandable as a gain factor that changes the slope of the f-i curve
 relating firing rate to input current.   
 If all of the neurons participating in the transform are modulated at each
 moment by $\alpha(t) = dx/dt$, then $F(s)$ holds the transform with respect
 to $x$ rather than time, $F(s) = \mathcal{L} f(x)$.  When one inverts the
 transform, with $\Lk$, this results in an estimate of the function of $x$,
 $\ftilde(\xstar) = \Lk F(s)$ (Figure~\ref{fig:CBB}A).  

 This strategy can be used to describe different kinds of functions by coding
 for different input stimuli---different ``what'' information---and
 choosing $\alpha(t)$ to be the rate of change of different variables. 
 In this section, we discuss computational work representing compressed
 functions of variables other than time.  For instance we will see that this
 approach can be used to compute functions coding for the relative spatial
 position of the wall of an enclosure, for past movements as a function of
 their position in the sequence or the amount of evidence accumulated for one
 of two alternatives in a simple decision-making task.
 The first subsection, entitled ``Spatiotemporal trajectories in the medial
 temporal lobe,'' reviews evidence that transform/inverse pairs of
 representations can account for a ``particle zoo'' of functional cell types
 in the MTL during spatiotemporal navigation.  In the next subsection,
 entitled  ``Accumulated evidence and decision-making,'' we describe a neural
 implementation of widely-used cognitive models for evidence accumulation
 models using transform/inverse pairs.  
 Finally, in the last subsection,
 entitled ``Cognitive models built entirely of transform/inverse pairs,'' we
 consider the possibility of cognitive models made entirely of
 transform/inverse pairs and how they could exploit computational properties
 of the Laplace domain for symbolic computation.

\nocite{FranEtal00,WoodEtal00}

 \subsection*{Spatiotemporal trajectories in the medial temporal lobe}

 \label{sec:spatiotemporal}
 It has long been suggested that the hippocampal place code is a special case
 of a more general form of representation coding for spatial, temporal and
 other more abstract relationships between events
 \cite{OKeeNade78,CoheEich93,EichEtal99,Hass12}. 
 A wide diversity of functional cell types that
 communicate information about kinematic variables have
 been reported in the hippocampus and related structures, including place
 cells, border cells, splitter cells, trajectory coding cells, speed cells,
 head direction cells and many more.  Many of these functional cell types
 (the most notable exception being grid cells) can be understood as the
 Laplace transform of a function coding a spatiotemporal trajectory; others
 can be understood as an approximate inverse of a function.  Moreover, these
 populations seem to come in pairs, with populations with properties like the
 Laplace transform in the entorhinal cortex and the populations with
 properties like the inverse in the hippocampus.

 Consider border cells in the medial entorhinal cortex (MEC)
 \cite{SolsEtal08}.  Border cells fire maximally at a location close to the
 boundary of an environment with a particular orientation.  Their firing rate
 decays monotonically with distance to the boundary.  We saw earlier that
 temporal context cells in the entorhinal cortex are perturbed by a specific
 stimulus and then relax monotonically towards their baseline firing rate over
 time (Fig.~\ref{fig:Timecells}C, \cite{TsaoEtal18,BrigEtal19}).   Temporal
 context cells code the Laplace transform of a function over time $F(s) =
 \mathcal{L} f(\tau)$.    The Laplace transform of distance to the boundary
 $F(s) = \mathcal{L}f(x)$ would behave similarly, with
 exponentially-decaying receptive fields in space.  As the animal moves away from a
 cell's preferred boundary, firing rate would decrease exponentially with
 distance; as the animal moved towards the boundary the firing rate would
 increase along the same curve describing the receptive field
 (Fig.~\ref{fig:CBB}B).  This is possible because $\alpha(t)$ is the signed
 velocity in the direction of the boundary.   If a population of border cells
 encodes the Laplace transform of distance to the boundaries, then  across
 neurons there should be a wide variety  spatial receptive field sizes, and
 more neurons should have narrow spatial receptive fields than wide spatial
 receptive fields.  The continuum of spatial locations should be mapped onto
 a continuum of values of $s$ in the population of border cells.
 
 By analogy to time cells, which have receptive fields in a circumscribed
 region of time since a triggering event, the inverse of border cells would
 generate a population of neurons with circumscribed receptive fields in
 space.  Boundary vector cells (BVCs), observed within the subiculum, have just this
 property, with elongated firing fields that align with boundaries of an
 enclosure \cite{LeveEtal09}.  In fact classical hippocampal place cells
 behave as if they are formed from conjunctions of BVCs 
 \cite{OKeeBurg96,BarrEtal06}.  If BVCs and place cells are
 the result of an approximate inverse transform, they should have
 properties analogous to those observed for populations of time cells.
 BVCs should have more fields close to boundaries and the width of fields
 should increase with distance to the boundary.   

 This framework organizes other ``cell types'' in the MTL as
 well.  Consider a population of cells coding for the sequence of movements
 leading up to the present position as a function of distance traveled.  In
 words, this population codes for a function $f$ that carries
 information like ``I got here by travelling North for 2~cm; before that I
 moved West for 10~cm \ldots''  In this case, the ``what information'' in the
 population would be head direction (``2~cm in the past I was facing North'' or 
 ``8~cm in the past I was facing West'').  In order to convey this information
 as a function of traveled distance, we would set  $\alpha(t)$ to be speed
 (unsigned velocity in the direction of motion).  Cells coding the Laplace
 transform of this kind of function would behave as ``trajectory-dependent''
 or ``retrospective coding'' cells \cite{FranEtal00}.  Cells coding for the
 inverse transform would manifest as ``splitter'' cells that fire
 differentially on the central arm of a figure-8 maze during an alternation
 task depending on the past locations \cite{FranEtal00,WoodEtal00,DudcWood14}
 that have been observed in the entorhinal cortex and hippocampus.

 Other functional cell types in the MEC can be understood as coding for
 spatiotemporal trajectories in the Laplace domain or approximating functions.
 When an animal pauses during a virtual navigation task, a population of MEC
 cells fire sequentially recording the amount of time since the animal ceased
 moving \cite{HeysDomb18}.  Even speed cells, which are believed to map the
 animal's instantaneous speed onto their firing rate, actually filter speed as
 a function of time with a spectrum of time constants \cite{DannEtal19}.  This
 is consistent with the idea that speed cells in MEC are actually coding the
 Laplace transform of the history of speed in the time leading up to the
 present.  The characteristic predictions from this theoretical approach are
 best evaluated at the level of populations and manifest largely as
 distributions of parameters.
%


 \subsection*{Accumulated evidence and decision-making}

 \begin{figure}
		\centering
       		\includegraphics[width=0.95\textwidth]{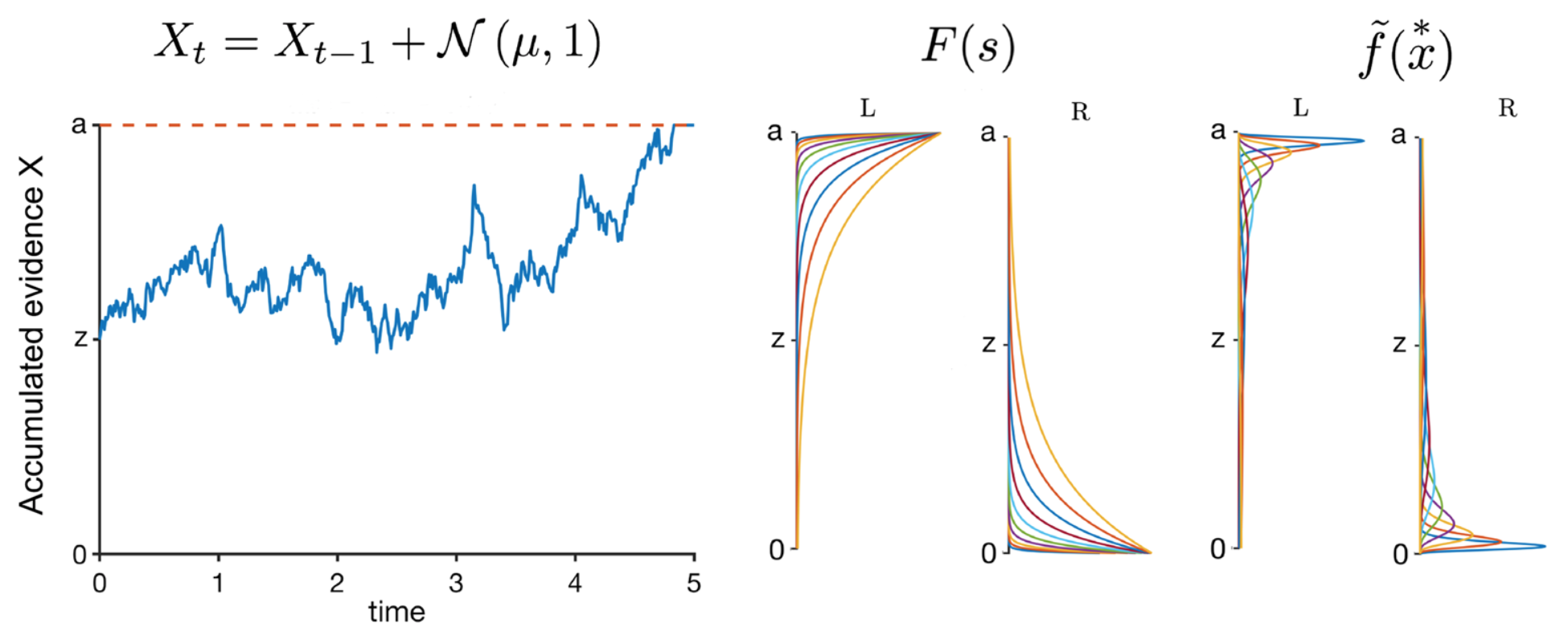}
		\caption{\textbf{A Laplace-domain implementation of the diffusion model
		for evidence accumulation.}  The diffusion model describes the internal
		state while a decision is being made as a particle moving towards two
		absorbing boundaries, each corresponding to one of two possible
		decisions.  At each moment of the decision, the position of the
		particle is a delta function located at a position $X_t$ that starts
		at a position $z$ and moves
		between the two boundaries.  In the Laplace domain implementation of
		the diffusion model, two populations code for the distance to each of
		the two decision bounds.   Cells coding for the Laplace transform of
		distance-to-bound, $F(s)$, will ramp their firing up (or down) as evidence
		accumulates.  These cells have exponential receptive fields over the
		decision axis.  Different populations code for each of the two
		boundaries.  We distinguish the two populations as $F_L(s)$ and
		$F_R(s)$.  Within each population, different neurons have different
		values of $s$.  Cells coding for the inverse Laplace transform,
		$\ftilde(\xstar)$, have receptive fields that tile each decision axis.
		Within each population different cells have different receptive field
		centers.  After Howard, et al., (2018).
		}
		\label{fig:EvidenceAccumulation}
 \end{figure}

\label{sec:evidenceaccumulation}
 In many simple decision-making experiments,  noisy instantaneous evidence
 must be integrated over time in order to reach a confident decision.
 Decades of work in mathematical psychology has resulted in sophisticated
 computational models for simple evidence accumulation tasks
 \cite{Luce86,SmitRatc04}.  The best known is the diffusion model
 \cite{Ratc78} (Figure~\ref{fig:EvidenceAccumulation}).  At each moment during
 the decision, the observer samples some evidence.  The ``particle's
 position'' at any moment, $X_t$, describes the accumulated evidence for each
 alternative up to that point.  This abstract model aligns to a strategy in
 which the starting position (usually referred to as $z$) is
 controlled by the decision-maker's prior expectations and the boundary
 separation (usually referred to as $a$) describes the degree of confidence
 the decision-maker requires before making a choice \cite{GoldShad07}.   
 We can understand the evidence at any moment $t$ as a function $f(x)$ with a peak
 at a single value $X_t$.  The time derivative of the position of the particle
 is just the instantaneous evidence sampled at time $t$. 

 With this understanding, it is straightforward to build a Laplace-domain
 model of the diffusion model by constructing two populations, one of which
 estimates the distance of $X_t$ to the lower bound and one that estimates the
 distance to the upper bound.  We will subscript the two populations
 such that $F_R(s)$ and $F_L(s)$ correspond to the Laplace transforms of these
 two functions and $\ftilde_R(\xstar)$ and $\ftilde_L(\xstar)$ correspond to
 the inverse transforms (Figure~\ref{fig:EvidenceAccumulation},
 \cite{HowaEtal18}).  In the diffusion model evidence for one alternative
 reduces the evidence for the other alternative so $\alpha_L(t) =
 -\alpha_R(t)$.  A decision is made when the ``particle'' reaches the smallest
 value of $\xstar$ in one of the populations.  Setting $\alpha_L$  and
 $\alpha_R$ to be non-zero and have the same sign effectively changes the
 decision bounds.  Positive paired values of $\alpha$ have the effect of
 widening the decision bounds whereas negative paired values of $\alpha$ have
 the effect of collapsing the decision bounds enabling speeded decisions
 \cite{PaleEtal18}.
 
 This Laplace-domain neural implementation of the diffusion model has at most
 subtle behavioral differences from the diffusion model.    However, it is
 quite distinct from other neural implementations of evidence accumulation.
 Rather than assuming that the abstract decision variable $x$ is carried by
 the average firing rate of many neurons \cite{ZandEtal14}, the Laplace domain
 representation represents $x$ as a distributed pattern of firing across many
 neurons indexed by their value of $s$.   This population has receptive fields
 that are exponential curves in the decision variable, much like leaky
 integrator models for decision-making \cite{BuseTown93}.  This approach
 further predicts that there should be a heterogeneous distribution of $s$
 values across neurons, analogous to recent findings from rodent cortex
 \cite{KoayEtal19}.  It is precisely this heterogeneity across neurons---the
 fact that $s$ forms a continuum---that allows the population to code the
 Laplace transform of accumulated evidence.  The inverse transform leads to
 neurons with compact receptive fields along the decision axis, analogous to
 empirical findings from rodent posterior parietal cortex \cite{MorcHarv16}.

\subsection*{Cognitive models built entirely of transform/inverse pairs}
 \label{sec:cognitivearchitecture}
 We have seen evidence that memory---data represented as functions over time
 and space---and evidence accumulation---a function over position within a
 decision space---can both be represented with the same form of neural circuit
 for encoding the Laplace transform and inverse.   Many detailed cognitive
 models of memory tasks include a memory component and an evidence
 accumulation component in describing behavioral data
 \cite{Ratc78,NosoPalm97,SedeEtal08,DonkNoso12}.  Moreover, detailed models of
 evidence accumulation make use of memory for past outcomes to make sense of
 sequential dependencies in RTs \cite{Korn73} manifest as changes in bias,
 drift rate or boundary separation \cite{GoldEtal08,UraiEtal19}.  Neurally,
 reward history can be decoded from neural activity in several brain regions
 \cite{MorcHarv16,MarcEtal13}.  Bernacchia and colleagues \cite{BernEtal11}
 estimated the time scale over which cortical neurons were modulated by reward
 history and found a wide range of decay rates, very much consistent with the
 idea that the population contained information about the Laplace transform of
 the history of rewards.  

 This convergence between memory and evidence accumulation suggests the
 possibility  that two interrelated systems using the same mathematical form
 interact with one another.  That is, perhaps the same equations that govern
 memory for reward history over tens of minutes \cite{BernEtal11}  also govern
 the  evolution of evidence between two decision bounds over the scale of less
 than a second \cite{KoayEtal19}.  The reward history could be used to set the
 bias parameter of the evidence accumulator so that segment of the
 computational cognitive model could be built from the same form of equations.
 Cognitive architectures \cite{Lair12,Ande13} have long provided
 self-contained models for cognitive performance, with many interacting
 modules contributing to any particular task. 
 Perhaps if the same neural circuit can be used for the evidence accumulation and
 working memory modules, one could use the same kind of canonical Laplace
 circuit to construct all (or most of) the modules needed to perform a
 complete task.  
 A general cognitive computer built along these lines would require sequential
 operation of ``cognitive programs'' operating on circuits representing
 information as functions.  Evidence accumulation circuits, along with the
 property that $\alpha(t)$ can be externally set, could be used to implement
 conditional flow control of the sequence of operations. 
 We discuss two additional considerations that suggest the Laplace domain
 could be well suited for a more general cognitive computer.  First, it is
 mathematically straightforward to write out efficient data-independent
 operators using the Laplace representation.  These can be understood as
 population-level modulations of circuits and, at least in the case of the
 translation operator---can lead to interesting connections to
 neurophysiology.  Second, neural evidence suggests neural representations of
 sequences of motor actions can be understood as functions over planned
 future time. This suggests that other sequences---for instance sequences of
 cognitive operations---could also be constructed as functions over a planned
 future.  These properties are necessary (but certainly not sufficient) to
 develop a general computing device to mimic human cognition
 \cite{GallKing11}.  
 Early computational work has demonstrated the feasibility of this approach
 at least for a few simple laboratory memory tasks \cite{TigaEtal19}.  

 \subsubsection*{Efficient data-independent operators in the Laplace domain}

 The properties of the Laplace domain make it particularly well-suited for
 data-independent operations.  Given data---in the form of functions
 represented as  $F(s)$/$\ftilde(\xstar)$ pairs---these operators generate
 an appropriate answer for every possible function they could encounter. 
 For instance, an addition operator should not need to know in advance what
 pair of numbers will be added together and should work effectively on numbers
 it has never experienced before.   Properties of the Laplace domain provide
 efficient recipes for data-independent operators.  We discuss several of
 these here.
 
 The translation operator takes a function
 $f(x)$ and shifts it by some amount
 to $f(x+\delta)$.  Consider how to implement translation of a function
 represented by a neural population $\ftilde(\xstar)$.  We would need to
 transfer information from each cell $\xstar_o$ to a translated cell $\xstar_o
 + \delta$.  This could be implemented \emph{via} a functional connection
 between pairs of cells---i.e., a matrix of synaptic connections.  However,
 because we do not know \emph{a priori} what displacement $\delta$ will be
 required, to be useful for all possible translations this hypothesized
 circuit must connect every neuron in $\ftilde(\xstar)$ with every other
 neuron. Translation in the Laplace domain is computationally
 more simple.  If  $F(s)$  is the transform of $f(x)$, the transform of the
 translated function $f(x+\delta)$ is simply $e^{-s\delta}F(s)$.  That is, the
 activity of each cell coding for the transform is multiplied by a number that
 depends on $s$ and $\delta$.  There is no need for information to be
 exchanged \emph{between} cells in $F$.  To examine the translated function,
 we simply need to invert the transform with $\Lk$ and obtain an estimate of
 $f(x+\delta)$.  

 Translation is potentially useful for many problems that arise in cognitive
 science.
 For instance, translating functions over time can be used to predict the
 future.  A model implementing function translation to predict the future
 \cite{ShanEtal16} can be mapped on to theta phase precession in the
 hippocampus and associated regions \cite{MeerEtal11}.   The key
 neurobiological property necessary to implement translation in this model is
 the ability to dynamically modulate synaptic weights over the course of theta
 oscillations, a property that has been observed in field potential recordings
 \cite{WyblEtal00}.  Translation could also be useful in manipulating visual
 representations or generating planned movements in allocentric space
 \cite<e.g.,>{JohnRedi07}.  
 

 Translation by a specific amount is a unary operation.  That is, translation
 takes in one piece of data---here a neural estimate of a function---and
 returns its answer.  The Laplace domain provides efficient recipes for other
 data-independent unary operators---for instance estimating the moments of a
 distribution or the derivative of a function.  A cognitive computing language
 would also requires binary operations.  For instance, addition is a binary
 operator---it requires two pieces of information to return its answer.   
 Consider what addition of two functions would mean computationally.  Suppose
 we have two functions $f(x)$ and $g(x)$ representing two specific numbers
 $x_f$ and $x_g$.  We can imagine $f(x)$ as a flat function except for a peak
 at the value $x_f$ and $g(x)$ as a flat except for a peak at $x_g$.  What
 would we desire for a function representing the sum of these two numbers?
 Simply adding $f(x) + g(x)$ is clearly not what we want---this would give two
 peaks, one at $x_f$ and the other at $x_g$, which is not understandable as a
 single number.  A moments reflection shows that we would want the
 representation  $[f + g][x]$to have a single peak at $x_f + x_g$.  The
 \emph{convolution} of two functions produces just this answer.  The
 convolution of two functions is written $f \star g$.  Much like translation,
 convolution performed directly on functions is computationally demanding.  To
 directly convolve a population of cells $\ftilde$ and another population
 $\tilde{g}$ would require one to take the product of the activation of all
 possible pairs of cells and then sum the results, keeping
 separate the information about the difference in $\xstar$ between them.
 While this is possible to compute it would require many connections and a
 relatively elaborate circuit.  In contrast, convolution is much more simple
 in the Laplace domain.  In particular, $\mathcal{L}\left[ f \star g \right] =
 F(s) G(s)$.  That is, to construct the transform of the convolution of two
 functions, we need only take the product of the transform of each of the
 functions at each $s$.  To invert the transform and obtain a direct estimate
 of the answer, we would apply $\Lk$ as $\Lk F(s) G(s)$.  A neural circuit
 implementing this mathematical operation would provide a sensible answer for
 any pair of numbers.

 To subtract a pair of numbers, we need an operator that is the inverse of
 addition.  The inverse to convolution is referred to as cross-correlation, 
 $f \# g$.  Like convolution, the Laplace transform of the cross-correlation
 of two functions is relatively simple: $\mathcal{L}\left[f \# g\right] = F(s)
 G(-s)$.  Although there are certainly important problems to solve in
 constructing a detailed neural model of subtraction, existence of an
 inverse operator to addition eliminates a conceptual obstacle to constructing
 a number system: any pair of ``numbers'' (represented as functions over a
 popuation of neurons) could be combined to obtain a new ``number.''  
 The compression of neural representations $\ftilde(\xstar)$  means that our
 estimate of number is not precise, but of course the brain's estimate of
 number is also imprecise \cite{GallGelm00,FeigEtal04,NiedDeha09}.    Notably,
 the quantitative form of compression of the brain's number system is believed
 to be similar to the compression of retinal coordinates in the cortex
 \cite{Schw77,VanEEtal84} and is at least roughly consistent with the form of
 compression of time shown by time cells.  Moreover, a general subtraction
 operator could be used across many different cognitive domains.  
 
 
%

 \subsubsection*{Are planned actions represented using the Laplace
 domain?}
 In order to build ``cognitive programs''  it would be necessary to
 sequentially gate information in and out of memory and to and from the
 evidence accumulation circuit.  Gating can in principle at least be
 implemented \emph{via} oscillatory dynamics in the brain
 \cite{SherEtal19,BhanBadr18}.  But this still leaves the question of 
 how the brain could construct plans for sequential application of various
 operations.   Although little is known about neurophysiological signatures of
 sequential plans for abstract computations, there is a good deal of
 neurophysiological evidence about plans for sequential plans for motor
 operations.  This literature is at least consistent with the idea that a
 these plans could be represented over planned future time.

 Sequential plans of motor actions can be understood as a function describing
 what \emph{will} happen when in the planned future.  In a task where monkeys
 had to make a series of movements, recordings from the lateral PFC 
 showed neurons that code for what motion an animal makes in sequence
 \cite{MushEtal06}.  That is, the animal had to perform a series of movements,
 say moving a cursor \textsc{left-right-down}.  As the sequence unfolded,
 cells fired conjunctively for specific movements (e.g., \textsc{left} or
 \textsc{down}) but only in specific positions in the sequence (e.g., first,
 second, or third).  In much the same way a stimulus-specific time cell fires
 only when its preferred stimulus is in its temporal receptive field
 \cite{TigaEtal18a,TaxiEtal18}, these cells fire when their preferred movement
 occurs in their sequential receptive field.  Notably, these populations in
 lPFC also fired in the moment before the sequence was initiated
 \cite{MushEtal06}, but retaining the coding properties that will occur in the
 future movement.   This pre-movement firing  was  as if the entire sequential
 plan was quickly loaded into memory prior to movement initiation.
 Smooth reaching movements also result in sequentially activated cells in
 motor cortex \cite{LebeEtal19}.  The similarity to stimulus-specific time
 cells suggests that these neural populations could code for an estimate of a
 function of sequential actions.  

 By analogy to the Laplace transform of the past, cells coding the Laplace
 transform of planned future actions would manifest as cells that ramp
 \emph{to} the time when an event will take place.  Ramping neurons during
 movement  preparation have been observed in prefrontal cortex 
 cortices \cite{Nara16},  including anterior lateral motor (ALM) cortex
 \cite{LiEtal16,InagEtal18,SvobLi18,InagEtal19}.   Neurons in ALM in
 particular can be used to decode what movement will occur and how far in the
 future \cite{LiEtal16}.  Note that when $s$ is small, an exponential function
 is approximately linear.  If this ALM population codes for the Laplace
 transform of time until a planned movement, this predicts that different
 cells should ramp at a variety of rates.

%
%

\section*{Discussion}

 This paper pursues the hypothesis that the brain represents functions in the
 world as activity  over populations of neurons.  The parameters of the
 receptive fields of these neurons trace out a continuum and the brain uses
 two distinct forms of receptive fields.  Exponential receptive fields enable
 a population to code for the Laplace transform of a function; circumscribed
 receptive fields enable a compressed estimate of the function itself.  We reviewed
 evidence that the brain maintains both of these kinds of representation for
 functions over past time in the EC and hippocampus.
 Computationally, this approach can be used to estimate
 functions over many other variables.  Considering spatial variables we can
 make sense of border cells, boundary vector cells  and other functional cell
 types in the hippocampus and related regions.  We reviewed computational work
 showing that widely-used cognitive models for evidence accumulation can be
 cast in this framework, making distinctive neural predictions.

 \subsection*{Computational neuroscience and cognition}
 As our ability to measure activity from large numbers of neurons grows, 
 it will be increasingly necessary to have ways of understanding the
 collective behavior of large numbers of neurons \cite{Yust15,Hass15}.  The
 basic unit of analysis we have argued for is not the neuron, but rather
 populations of neurons representing and manipulating continua. This is
 analogous to the approach taken in many fields of physics where it has long
 been appreciated that theories should describe phenomena at an appropriate
 level of detail \cite{Ande72a}.  For instance, fluid dynamics describes the
 flow of liquids not in terms of molecules but in terms of incompressible
 volume elements.   To determine the flow of water in a pipe one does not need
 need to worry at all about chemistry.  If we could measure the position of
 each individual water molecule during an experiment, we could evaluate the
 theory, but the theory would be equally correct no matter whether we
 understand the chemistry of water molecules or if the incompressible volume
 element was made of green cheese.  A different theory would be required to
 understand why some liquids have different viscosity than others.  Returning
 to neuroscience, if the approach in this paper has merit, it suggests a
 number of specific problems that are tractable in the context of
 computational neuroscience.
 How does a population of temporal context cells manage to have a specific
 distribution of time constants?   How do neural circuits implement $\Lk$?  We
 discuss some possibilities in Appendix~1, but the larger point is that this
 approach segments the computational neuroscience of circuits of neurons from
 cognitive neuroscience.  Understanding cognition starting
 from individual neurons is kind of like trying to understand the flow of
 water through a channel starting with a model of the Bohr atom.
 
 If it is really the case that populations of neurons organize themselves to
 estimate continua, then this places constraints on the data analysis tools we  use
 to study populations of neurons.  Thus far, the strategy taken with time
 cells and temporal context cells has been to construct a hypothesis about the
 specific variable being represented and then estimate individual receptive
 fields to hopefully trace out a continuum of parameters across neurons
 corresponding to $\taustar$  or $s$.  This approach could in principle be
 applied piecemeal to problems in different brain regions, but there are
 significant challenges.  First, even in the hippocampus, cells have receptive
 fields along more than one kind of variable.  For instance,
 consider the situation when an animal is placed on a treadmill with varying
 speeds.  Because the speed changes from trial to trial, the time since the
 run started is deconfounded from distance traveled.  We would expect ``time
 cells'' to care only about time and ``distance cells'' to care about
 distance.  However, all possible combinations of time and distance
 sensitivity are observed, with time cells and distance cells as special cases
 of a continuous mixture \cite{KrauEtal13,HowaEich15}.  Second, our \emph{a
 priori} hypotheses about what a specific population of cells codes for depend
 on prior work.  A data-driven approach to neural data analysis would avoid
 these kinds of problems.  However, widely used data-driven approaches can be
 ill-suited to discover continua.   For instance, individual reaching
 movements generate sequences of activity in motor cortex strikingly similar
 to sequences of time cells \cite{LebeEtal19}.  These sequences can be readily
 understood as cells tiling a continuum, $\ftilde(\xstar)$.  But data-driven
 dimensionality reduction methods can identify rotational dynamics from the
 same kinds of data \cite{ChurEtal12,AoiEtal19}.  To overcome this problem
 will require data-driven tools that look for multidimensional continua in
 neural coding.   

 \subsection*{Computational models of natural and artificial cognition}
 
 Laboratory cognitive tasks allow us to study behavior in a quantitative way
 under tightly controlled circumstances.  Although this approach is quite
 artificial relative to real-world cognition, it  places strong constraints on
 computational models of behavior.  However, recent work in mathematical
 psychology has shown that even very successful cognitive models cannot be
 uniquely identified using behavioral data alone \cite{JoneDzha14}.   Joint
 modeling of neural and behavioral data is a promising avenue to constrain
 cognitive models \cite{TurnEtal13,TurnEtal17,PaleEtal18a}, but it does not
 solve the problem of determining whether a particular cognitive model is
 neurally plausible \emph{a priori}.  
 If we knew with certainty that populations of neurons really do represent
 continua \emph{via} the Laplace transform and that those continua have a
 specific form of compression, this would place a strong constraint on
 detailed cognitive models of behavior.  

 If thoughts map onto functions, then thinking maps onto manipulating those
 functions.  The Laplace domain provides recipes for data-independent
 operators that could be used to manipulate and compare
 functions---to think.  As such, this way of viewing cognition and
 neurophysiology sidesteps many of the conceptual concerns that have
 traditionally dogged connectionist models and much of contemporary deep
 learning approaches.  
 

 \clearpage 
\section*{Appendix~1: Possible neurophysiological mechanisms for the Laplace
transform}

 There are three main requirements to implement the Laplace transform/inverse
 coding scheme for functions over arbitrary variables.  First, the the Laplace
 transform requires that neurons have a wide range of functional time
 constants that are very large compared to  membrane time constants.  Second,
 to invert the Laplace transform it is necessary for a circuit to implement
 the $\Lk$ operator.  Third, to enable coding of the Laplace transform of
 functions other than time, it is necessary to manipulate the gain of neurons.
 This box sketches possible neurophysiological mechanisms for these three
 computational functions.  There are almost certainly other possible
 mechanisms that could give rise to these properties and there is no
 guarrantee that, even assuming that different brain regions obey the same
 equations, that they are implemented using the same mechanisms in different
 regions.  

 Neurophysiological data indicates that neural circuits could implement the
 mechanisms of the Laplace transform. The real part of the Laplace transform
 corresponds to exponential decay with a spectrum of time constants. 
 Recurrent network connections could generate slow time constants, but it is
 also possible that intracellular mechanisms contribute to exponential decay
 with a variety of time constants across cells.  Intracellular recordings in
 cortical slice preparations show persistent firing over a range of time
 scales in the absence of synaptic input.  For instance, spike frequency
 accommodation of neurons in piriform cortex shows a pattern of exponential
 decay over hundreds of milliseconds \cite{BarkHass94}.  In entorhinal cortex
 slice preparations show exponential decay in persistent firing rate over
 seconds \cite{TahvEtal07,KnauEtal13}.  At the upper limit, isolated  neurons
 in slices from entorhinal \cite{EgorEtal02} and perirhinal cortex
 \cite{NavaEtal11} integrate their inputs and maintain persistent firing for
 arbitrarily long periods of time.  These cells show effectively infinite time
 constants in the absence of synaptic inputs.  The decay in persistent firing
 can be modeled based on the properties of nonspecific calcium dependent
 cation current and calcium diffusion \cite{FranEtal06,TigaEtal15}. 

 The entorhinal cortex provides input to the hippocampus, so that the
 population coding for $F(s)$ is one synapse away from the population coding
 for $\ftilde(\taustar)$.  Because the equation $\ftilde(\taustar) = \Lk F(s)$
 is mathematically true, there should be some way to understand the functional
 mapping between the regions as an approximate inverse Laplace transform
 $\Lk$.   The inverse Laplace transform requires combining the different exponential
 decay rates with different positive and negative values
 (Eq.~\ref{eq:inverse}, Appendix~2). The simplest way to think of this is subtraction of an exponential
 function with a faster decay from an exponential of the same starting value
 with slower decay. This will result in a function that peaks at a time point
 dependent upon the difference of the two time constants. If we multiplied
 both of the time constants by the same number, the difference would peak at a
 proportionally larger time. A biological
 detailed spiking model of the inverse Laplace transform \cite{LiuEtal19} can
 be built from a series of additions and subtractions in which a particular
 time constant has subtractions from time constants close in value.   These
 derivatives with respect to $s$ (Eq.~\ref{eq:inverse}) are analogous to
 center-surround receptive fields \cite{MarrHild80}, only in $s$ rather than
 in retinal position.   Methods for blind source separation including
 independent component analysis and sparse coding have been shown to give rise
 to center-surround receptive fields in models of the visual system
 \cite{BellSejn97,OlshFiel96a}.   Higher order derivatives could be
 implemented by placing center-surround circuits in series.

 Generalizing the Laplace transform to dimensions other than time require a
 coordination of gain control of decay by factors such as running velocity to
 code spatial location.  In at least some studies, spatial attention works to
 enhance the gain of receptive fields \cite{McAdMaun99,MaunTreu06}.   In slice
 preparations, gain can be controlled by the variability of synaptic inputs
 \cite{ChanEtal02} and computational studies have suggested a number of
 possible mechanisms \cite{Silv10}, including active dendritic computation
 \cite{MehaEtal05,PoirEtal03}, and neuromodulatory agents such as
 acetylcholine \cite{BarkHass94,FranEtal06}.  

 \clearpage
\section*{Appendix~2: Mathematics of the Laplace transform}

 Consider a population of leaky integrators indexed by their rate parameter
 $s$.  Each of the neurons in this population receive the same input $f(t)$ at
 each moment and update their firing rate as
\begin{equation}
		\frac{dF(s)}{dt} =  -s F(s) + f(t).
		\label{eq:noalpha}
\end{equation}
 We understand $F(s)$ as describing the activity of a large number of neurons
 with many values of $s$.  Note that Eq.~\ref{eq:noalpha}  only requires
 information about $f$ and $F$ at the present moment.  However, the solution
 to Eq.~\ref{eq:noalpha} gives the real Laplace transform of the entire
 function $f(\tau)$ running backwards from the present infinitely far in the
 past:
\begin{equation}
		F(s) = \int_0^{\infty} f(\tau) e^{-s\tau} d\tau
		\label{eq:Laplace}
\end{equation}
 where we understand $f(\tau)$ on the right hand side to be the series of
 inputs ordered from the present towards the past.  That is, the $f(\tau)$ on
 the right hand side of Eq.~\ref{eq:Laplace} is related to $f(t)$ in
 Eq.~\ref{eq:noalpha} as $f(\tau) \equiv f(t-\tau)$.  Another way to say this
 is that $F(s)$, the pattern of activity  at time $t$, is the Laplace
 transform of the entire history of $f$ leading up to the present.

The Post approximation \cite{Post30} provides a neurally-realistic method for
approximately inverting the Laplace transform.  This method allows us to take
the set of cells coding for $F(s)$, each with a different value of $s$ and map
them onto a new population of cells that estimate the original function.  We
index those cells by a parameter $\taustar$ and write $\ftilde(\taustar)$ to
refer to the firing rate of the entire population.  The approximation of
$f(\tau)$ is computed as follows:
\begin{eqnarray}
		\ftilde(\taustar) & = &  \Lk F(s)\\
						  &= & C_k s^{k+1} \frac{d^k}{ds^k} F(s)
		\label{eq:inverse}
\end{eqnarray}
The parameter $k$ controls the
precision of the approximation.
Post proved that in the limit as $k \rightarrow \infty$, $\ftilde(\taustar)
= f(\tau)$.
The internal estimate of past time $\taustar$ is related to $s$ as 
$\taustar = k/s$.  This means that $\taustar$ is proportional to the time
constant $1/s$.    The value of $\taustar$ has a physical meaning in that it
gives the time lag at which each cell in $\ftilde$ would peak following a
delta function input.

Equation~\ref{eq:inverse} describes a mapping from a population of cells
indexed by $s$ to another population indexed by $\taustar$.  
To understand the
mechanism of the inverse operator, let's consider Eq.~\ref{eq:inverse} from the
perspective of a particular cell with a particular value $\taustar_o$.  
The time dependence on the right hand side comes entirely from the derivative
term---$C_k$ is a constant that is the same for all cells and $s^{k+1}$ is a
scaling factor specific to the value of $\taustar_o$.
The derivative term says that the firing rate 
$\ftilde(\taustar_o)$ is controlled by the $k$th derivative with respect to
$s$ in the neighborhood of a specific value of $s$,  $s_o = k/\taustar_o$.  
Computing the $k$th derivative requires comparing the firing rate other cells
in the neighborhood of $s_o$ \cite{ShanHowa13}.

 To generalize to functions over variables other than time, we allow the gain
 of all of the neurons coding for $F(s)$ to be modulated together by a
 time-dependent function $\alpha(t)$:
 \begin{equation}
		\frac{dF(s)}{dt} = \alpha(t) \left[ -s F(s) + f(t) \right]
		\label{eq:alpha}
 \end{equation}
 Note that if $\alpha(t) = 1$, this expression reduces to
 Eq.~\ref{eq:noalpha}.  Consider the situation where $f(t)$ is a delta
 function input at $t=0$.  This initializes the activation at~1 for all units.
 This is the Laplace transform of a delta function at $x=0$.   If, in the time
 after $t>0$, we find $\alpha(t) = 1$, $F(s)$ at time $t$ will code for the
 Laplace transform of the time since the delta function, $F(s) = e^{-st}$.
 However, if $\alpha(t)$ was some positive constant that was greater or less
 than one, $\alpha_o$ we would find $F(s) = e^{-s \left(\alpha_o  t\right)}$.
 That is, changing $\alpha_o$ from 1 is equivalent to making time go faster or
 slower.   If we found $F(s)$ in a state where $F(s) = e^{-sx}$  for some
 value of $x$, then we set $\alpha$ to some specific value $\alpha_o$, we
 would find after some time displacement $\Delta_t$ that $F(s)$ is now 
 \[
		e^{-s \alpha_o \Delta_t}{e^{-sx}} = 
			e^{-s \left(x + \alpha_o \Delta_t \right)}.
 \]
 If we could arrange for $\alpha_o$ to be the rate of change of $x$ during
 this interval, then our new value of $F(s) = e^{-s \left(x + \Delta_x
 \right)}$.  Note that this is true whether $\Delta_x$ is positive or
 negative.  This means that  during an interval where $f(t)=0$,  if $\alpha(t)
 = dx/dt$ then $F(s)$ records  the Laplace transform of $f(x)$ rather than
 $f(t)$.  

Although there are many variables that could be productively represented in
this way, there are two potentially important limitations to this approach.
First, if $f(t)$ is to be non-zero, $f(t)$ must be an implicit function of
$x$, $f[x(t)]$.  This makes sense if $f(t)$ corresponds to, say, contact with
a landmark in a spatial environment but can lead to complications in general.
Second, significant problems arise when one attempts to use this approach to
represent values of $x < 0$.  
To make this concrete, note that Eq.~\ref{eq:alpha} works for both positive
and negative rates of change---as we would expect in a spatial navigation task
where the animal can move either to the left or to the right.  Suppose one
starts with $F(s) = e^{-s0}$; each cell at a high firing rate.  If we set
$\alpha$ to be $\alpha_o$ and evolve Eq.~\ref{eq:alpha} for some time we find
$F(s) = e^{-s \alpha_o \Delta_t}$.  If $\alpha_o$ is positive, each of the
cells decay from $1$.  However, if $\alpha_o$ is negative, the cells increase
their firing rate exponentially from 1, growing without bound.  Moreover, the
inverse  operator does not not behave well as $x$ passes through zero.

\section{Acknowledgments}
 Alexander Howard helped with Figure~2.  Supported by ONR MURI
 N00014-16-1-2832 and NIBIB R01EB022864. 
 The authors thank Randy Gallistel, Per Sederberg, Josh
 Gold, Aude Oliva, and Nathaniel Daw for helpful conversations.  

\bibliography{/Users/marc/doc/bibdesk}
\end{document}

\clearpage
\section*{Oustanding Questions}
 \begin{itemize}
	\item How widely used is the Laplace transform in the brain?  We have seen
			that the brain codes for the Laplace transform of the past in EC
			and it is mathematically possible to compute Laplace transforms of
			functions of many other variables as well.  Despite some
			suggestive evidence, it is not known if populations of neurons use
			this form of representation for other continuous variables.  The
			signature of the real Laplace transform is that exponential
			receptive fields $e^{-sx}$ should have a continuum of parameters
			$s$ across neurons.  Compression implies that $s$ is not sampled
			uniformly over the population.
	\item Can the 1-D theory be generalized to arbitrary N-D coordinate
			systems?  Thus far the theory is worked out pretty well for
			functions over the interval from 0 to $\infty$.  In situations
			where the space is bounded, such as evidence accumulation one can
			handle these situations by tracking distance to each of the
			bounds.  But a general theory of computation in the brain would
			require generalization to variables that cover the entire plane,
			and perhaps coordinate systems with more complex topology.  
	\item Is it possible to  understand grid cells in the same
			formalism that gives rise to all of these other functional cell
			types?  Although the Laplace/inverse approach makes sense of most
			of the types of cells'' in the hippocampus and related regions,
			grid cells are a notable exception.   One possible avenue is that
			the periodicity of grid cells is somehow a consequence of
			including complex Laplace domain variables.
	\item Is it possible to actually engineer neurally realistic models for
			symbolic computing using the Laplace domain?  Although this
			theoretical approach avoids many of the conceptual problems in
			symbolic computing in connectionist systems, there are doubtless
			many important practical obstacles to overcome.  Some of these
			problems will certainly require additional theoretical work.
 \end{itemize}

\clearpage
\section*{Highlights}
 \begin{itemize}
	\item Recent evidence suggests that the past is represented in the
			entorhinal cortex as the Laplace transform.
	\item This theoretical framework can be used to represent functions over
			many different continuous variables.
	\item These representations can be used to build relatively high-level
			cognitive models for memory and decision-making.
	\item The Laplace transform is useful for data-independent operations,
			which are necessary for cognitive computation. 
 \end{itemize}

%% file: tcmmacros.tex
\newcommand{\myvec}[1]{\ensuremath{\mathbf{#1} } }
\newcommand{\mat}[1]{\ensuremath{\mathbf{#1}}}
\newcommand{\Mlim}{\ensuremath{\mathbf{M}_{\rm{lim} } } }
\newcommand{\Mc}{\ensuremath{\mathbf{M^{o }} }}
\newcommand{\Mbar}{\ensuremath{\overline{\mathbf{M } } } }
\newcommand{\Hbar}{\ensuremath{\overline{\mathbf{H } } } }
\newcommand{\Sbar}{\ensuremath{\overline{\mathbf{S } } } }
\newcommand{\Mpre}{\ensuremath{\mathbf{M}_{\rm{pre} } } }

\newcommand{\IN}{\textrm{{\tiny IN}}}

\newcommand{\tin}[1]{\ensuremath{\myvec{t}^{\IN}_{#1 } } }
\newcommand{\cin}[1]{\ensuremath{\myvec{c}^{\IN}_{#1 } } }
\newcommand{\hin}[1]{\ensuremath{\myvec{h}^{\IN}_{#1 } } }
\newcommand{\ti}[1]{\ensuremath{\myvec{t}_{#1 } } }
\newcommand{\T}[1]{\ensuremath{\myvec{T}_{#1 } } }
\newcommand{\Psymb}[1]{\ensuremath{\myvec{P}_{#1 } } }
\newcommand{\tc}[1]{\ensuremath{\myvec{t}^{o}_{#1 } } } 
\newcommand{\h}[1]{\ensuremath{\myvec{h}_{#1 } } } 
\newcommand{\s}[1]{\ensuremath{\myvec{s}_{#1 } } } 
\newcommand{\F}[1]{\ensuremath{\myvec{f}_{#1 } } }
\newcommand{\ci}[1]{\ensuremath{\myvec{c}_{#1 }} }
\newcommand{\K}[1]{\ensuremath{\myvec{k}_{#1} } }

\newcommand{\tavg}[1]{\myvec{\underline{t}}_{#1} }
\newcommand{\proj}[1]{\ensuremath{\mat{P}_{#1} } }
\newcommand{\projcomp}[1]{\ensuremath{\tilde{\mat{P}}_{#1} } }
\newcommand{\Mtf}{\ensuremath{\mat{M}^{{\textrm{{\tiny TF}}}}}}
\newcommand{\Mft}{\ensuremath{\mat{C}}}
\newcommand{\MA}{\ensuremath{\mat{M}_{A} } }
\newcommand{\dM}{\ensuremath{\mat{\delta M} } }
\newcommand{\dS}{\ensuremath{\mat{\delta S} } }
\newcommand{\dMc}{\ensuremath{\mat{\delta \Mc} }}
\newcommand{\dMbar}{\ensuremath{\overline{\dM}}}
\newcommand{\dSbar}{\ensuremath{\overline{\dS}}}  
\newcommand{\dMcbar}{\ensuremath{\overline{\dM}^{o}}} 
\newcommand{\Tm}{\ensuremath{\mat{T}_{\rm{model}}}}

\newcommand{\rankiness}{\ensuremath{\mathfrak{R}}}

\newcommand{\bra}[1]{\ensuremath{\langle#1|}}

\newcommand{\ket}[1]{\ensuremath{|#1\rangle}}

\newcommand{\braket}[2]{\ensuremath{\left\langle#1|#2\right\rangle}}

\newcommand{\ketbra}[2]{\ensuremath{\left|#1\left\rangle\right\langle#2\right|}}

\newcommand{\aold}{\ensuremath{\alpha_{O}}}

\newcommand{\assoc}[2]{\ensuremath{{#1}\rightarrow{#2}}}
\newcommand{\scassoc}[2]{\textsc{#1}\ensuremath{\rightarrow}\textsc{#2}}
\newcommand{\INCM}{\ensuremath{I_{NCM}}}
\newcommand{\anew}{\ensuremath{\alpha_{N}}}
\renewcommand{\vec}[1]{\ensuremath{\mathbf{#1}}}
\newcommand{\vel}[1]{\ensuremath{\vec{v}^{#1}}}
\newcommand{\tvec}[1]{\ensuremath{\vec{t}^{#1}}}
\newcommand{\that}[1]{%
\ensuremath{%
		\hat{\vec{t}}^{#1}
}%
}\newcommand{\tdot}[2]{%
\ensuremath{%
		\vec{t}^{#1}\cdot\vec{t}^{#2}
}%
}
\newcommand{\tsub}[1]{%
\ensuremath{%
		\vec{t}_{#1}
}%
}
\newcommand{\tIN}[1]{%
\ensuremath{%
		\vec{t}^{\IN}_{#1}
}%
}
\newcommand{\cIN}[1]{%
\ensuremath{%
		\vec{c}^{IN}_{#1}
}%
}
\newcommand{\csub}[1]{%
\ensuremath{%
		\vec{c}_{#1}
}%
}
\newcommand{\hIN}[1]{%
\ensuremath{%
		\vec{h}^{IN}_{#1}
}%
}
\newcommand{\hsub}[1]{%
\ensuremath{%
		\vec{h}_{#1}
}%
}
\newcommand{\fsub}[1]{%
\ensuremath{%
		\vec{f}_{#1}
}%
}
\newcommand{\fIN}[1]{%
\ensuremath{%
		\vec{f}^{IN}_{#1}
}%
}

\newcommand{\set}[1]{%
	\ensuremath{%
	\mathfrak{#1}
	}%
}
\newcommand{\none}{
	\nonumber \\
}
\renewcommand{\vec}[1]{
\ensuremath{%
	\mathbf{#1}
}%
}

\newcommand{\MTF}{\ensuremath{\mat{M}^{\textrm{\tiny TF}}}}
\newcommand{\MFT}{\ensuremath{\mat{M}^{\textrm{\tiny FT}}}}

\newcommand{\myinsfig}[1]{%
	\refstepcounter{figure}	\label{myfig:#1} \addtocounter{figure}{-1}
	\ifapamode{	
		\includegraphics[width=0.9\textwidth]{figs/#1.eps}
	}{			
		\includegraphics[width=0.4\textwidth]{figs/#1.eps}
	}{			
		\includegraphics[width=0.6\textwidth]{figs/#1.eps}
	}
}
\newcommand{\myinssmfig}[1]{%
	\refstepcounter{figure}	\label{myfig:#1} \addtocounter{figure}{-1}
	\ifapamode{	
		\includegraphics[width=0.4\textwidth]{figs/#1.ps}
	}{			
		\includegraphics[width=0.15\textwidth]{figs/#1.ps}
	}{			
		\includegraphics[width=0.4\textwidth]{figs/#1.ps}
	}
}

\newcommand{\myinsrotfig}[1]{ 
	\refstepcounter{figure}	\label{myfig:#1} \addtocounter{figure}{-1}
	\ifapamode{ 
		\includegraphics[height=0.9\textheight]{figs/#1.eps}
	}{	
		~\hspace{0.05\textwidth}
		\includegraphics[angle=270,width=0.4\textwidth]{figs/#1.eps}
	}{ 	
		\includegraphics[angle=270,width=0.8\textwidth]{figs/#1.eps}
	}
}
\newcommand{\myinsbrotfig}[1]{ 
	\refstepcounter{figure}	\label{myfig:#1} \addtocounter{figure}{-1}
	\ifapamodeman{
		~\hspace{0.15\textwidth}
		\includegraphics[height=0.9\textheight]{figs/#1.ps}
	}{
		~\hspace{0.1\textwidth}
		\includegraphics[angle=270,width=0.8\textwidth]{figs/#1.ps}
	}
}

\newcommand{\Edit}[1]{
	{\large{\bf#1}}
}

\newcommand{\pair}[2]{\textsc{#1}-\textsc{#2}}

\newcommand{\Lk}{\ensuremath{\mathbf{L}^{\textrm{\scriptsize{-1}}}_{\textrm{k}}}}
\newcommand{\Tau}{\ensuremath{\overset{*}{\tau}}}
\newcommand{\taustar}{\ensuremath{\overset{*}{\tau}}}
\newcommand{\xstar}{\ensuremath{\overset{*}{x}}}
\newcommand{\pstar}{\ensuremath{\overset{*}{p}}}
\newcommand{\lstar}{\ensuremath{\overset{*}{p}}}
\newcommand{\Tauf}{\ensuremath{\overset{o}{\tau}}}
\newcommand{\po}{\mathcal{P}}
\newcommand{\To}{\mathcal{T}}
\newcommand{\Io}{\mathcal{I}}
\newcommand{\Ro}{\mathcal{R}}
\newcommand{\ftilde}{\ensuremath{\tilde{f}}}